\documentclass[smallcondensed]{svjour3}    
\usepackage{graphicx}
\usepackage{bm}
\usepackage{natbib} 
\usepackage{booktabs}
\usepackage{amssymb}
\usepackage{color}
\begin{document}

\title{Detection of Earth-mass and Super-Earth Trojan Planets 
Using Transit Timing Variation Method}

\titlerunning{Detecting Trojan Planets with TTV} 

\author{Nader Haghighipour         \\
        Stephanie Capen            \\
        Tobias C. Hinse
}

\authorrunning{Haghighipour, Capen \& Hinse} 

\institute{N. Haghighipour \at
 Institute for Astronomy and NASA Astrobiology Institute, University of Hawai'i-Manoa, 2680 Woodlawn Drive,
 Honolulu, HI 96822, USA, Tel.: +1-808-956-6098, Fax: +1-808-956-4351, \email{nader@ifa.hawaii.edu} \\
S. Capen \at
Department of Education, University of Hawai'i-Manoa, Honolulu, HI 96822, USA, 
\email{stephaniecapen@gmail.com.}
 \\
T. C. Hinse \at
Korea Astronomy and Space Science Institute, 304-358 Daejeon, Republic of Korea,
and Armagh Observatory, College Hill, BT61 9DG, Armagh, UK
\email{tchinse@gmail.com}
}        
\maketitle

\begin{abstract}
We have carried out an extensive study of the possibility of the detection of Earth-mass
and super-Earth Trojan planets using transit timing variation method with the {\it Kepler}
space telescope. 
We have considered a system consisting of a transiting Jovian-type planet in a short period orbit, and 
determined the induced variations in its transit timing due to an Earth-mass/super-Earth  
Trojan planet. We mapped a large section of the phase space around the 1:1 mean-motion resonance
and identified regions corresponding to several other mean-motion resonances where the orbit of the planet
would be stable. We calculated TTVs for 
different values of the mass and orbital elements of the transiting and perturbing bodies as well as 
the mass of central star, and identified orbital configurations of these objects
(ranges of their orbital elements and masses) for which the resulted TTVs 
would be within the range of the variations of the 
transit timing of {\it Kepler}'s planetary candidates. Results of our study indicate that in general, 
the amplitudes of the TTVs fall within the detectable range of timing precision obtained from 
the {\it Kepler}'s long-cadence data, 
and depending on the parameters of the system, their magnitudes may become as large as a few hours. 
The probability of detection is higher for super-Earth Trojans with slightly eccentric orbits around 
short-period Jovian-type planets with masses slightly smaller than Jupiter. We present the details 
of our study and discuss the implications of its results.


\end{abstract}

\section{Introduction}

It is generally accepted that similar to the giant planets of our solar system, Jovian-type exoplanets 
can host satellites and may have co-orbital and/or Trojan bodies. During the past decade, these considerations
have led to initiatives on both modeling the formation of these objects (especially in system with close-in 
giant planets) and assessing their detectability in particular using ground-based observational facilities. 
The success of the {\it Kepler} space telescope in detecting several planets using transit 
timing variation (TTV) method, combined with the unprecedented photometric sensitivity of this telescope 
and its high observing cadence has raised the question that whether Trojan planets 
and satellites of giant exoplanets can also be detected by the {\it Kepler}. In this paper, 
we address this question and study the prospects of the detection of these objects by comparing their induced 
transit timing variations with the transit timing precision 
of planetary candidates that have been identified by the {\it Kepler} space telescope.

Since the discovery of the first extrasolar planet,
many efforts have been made to develop models for the detection of the satellites of 
these objects. Sartoretti and Schneider (1999) were the first 
to study the possibility of the detection of exomoons using transit photometry. Deeg (2002) extended
this analysis to smaller objects and examined the photometric detectability of terrestrial-size satellites. 
Subsequently, Simon et al (2007) studied the transit timing variations induced by exo-satellites as
a mechanism for the indirect detection of these bodies. More recently, in a series of articles,
Kipping has shown that the measurements of the variations in the time as well as the duration of the transit 
will allow for the detection and determining the mass of the exomoon of a transiting giant planet 
(Kipping 2009a\&b, 2011).

Efforts have also been made to determine the possibility of the detection of co-orbital and Trojan planets. 
It has been shown by Laughlin and Chambers (2002) that a pair of Jupiter-like planets in a 1:1 mean-motion
resonance (MMR) in a system similar to that of GJ 876 can generate stellar radial velocities with 
amplitudes ranging from 50 m/s to over 100 m/s. These results agree with the more recent findings
of Giuppone et al (2012) who studied the origin and RV-detectability of co-orbital planets.
Laughlin and Chambers (2002) showed that as this range of radial velocity
falls within the sensitivity of ground-based telescopes, 1:1 resonant planets can be detected using
the Doppler Velocimetry technique when the fitting routine takes into account the 
mutual interactions of these objects.

The detectability of Trojan planets have also been studied by Ford and Gaudi (2006). These authors
showed that the perturbation of a planet in the ${\rm L}_4$ or ${\rm L}_5$ Lagrangian points of a
transiting Jovian-type body can cause the time when the stellar RV coincides 
with the RV of the barycenter of the star-transiting planet system to differ from the time of
the planet's mid-transit. As a result, a Trojan planet can create a time offset between the
ephemerides obtained from RV and transit photometry. As shown by these authors, 
these variations present a potential pathway for detecting terrestrial-mass Trojans
using ground-based telescopes. In a subsequent article, Ford and Holman (2007) examined the sensitivity
of transit timing observations as a mechanism for detecting Trojan planets, and showed that the 
transit timing variations induced by a terrestrial-class Trojan may fall within the range of the photometric
sensitivity of ground-based observational facilities.

The above-mentioned vast interest in the discovery of Trojan planets has roots in the fact that 
while the theories of planetary dynamics support the existence of these objects, no Trojan planet exists in
our solar system. As such, the detection of these bodies will be a novelty that will have profound 
effects on the 
models of planet formation and dynamics. Also, given that many giant planets reside in the
habitable zones of their host stars, the Trojans (and satellites) of these objects, if existed 
and of terrestrial-size, could be potentially habitable. While these characteristics of Trojan planets
make them interesting subjects for observational surveys, and although previous studies
suggest that these objects are detectable using the current ground-based observing facilities,
no Trojan planet has yet been found. The latter may be the indication of the inefficiency of
ground-based surveys in searching for this type of planets. The space-based telescopes, especially 
{\it Kepler}, on the other hand, have been able to detect many planetary bodies 
of different sizes using transit photometry and transit timing variation method 
(Holman et al 2010; Nesvorn\'y et al 2012). 
From these two techniques, TTV is particularly efficient in detecting
Earth-mass and super-Earth planets, especially when these objects are in mean-motion resonance with their
corresponding transiting bodies (Agol et al 2005, 2007; Holman and Murray 2005; Steffen and Agol 2005;
Heyl and Gladman 2007; Haghighipour and Kirste 2011; Veras et al 2011). The latter makes the 
foundation of our study. We consider systems with transiting giant planets in short-period orbits, 
and study the prospects of the detection of putative Earth-mass and super-Earth Trojans 
using transit timing variations method with the {\it Kepler} space telescope.

Our approach is numerical and based on calculating TTVs and comparing their amplitudes with 
the amplitudes of the variations in the transit
timing of the {\it Kepler}'s planetary candidates as reported by Ford et al (2011).
We begin by studying the stability of our Trojan planets and identifying the regions of the
phase space, in particular resonances around 1:1 MMR, where the orbit of the planet is stable. 
We then calculate TTVs for the 1:1 MMR configurations, assuming that an Earth-mass or super-Earth 
Trojan exists in those stable regions. 
In this paper, we will not be concerned with the formation of Trojans and how our Trojan 
planets acquired their orbits. Instead, we will focus our study on merely identifying regions of 
the parameter space
for which the amplitudes of TTVs fall within the range of the TTVs of the planetary candidates
detected by the {\it Kepler}.
We will also not intend to break the inherent degeneracy associated with
the TTV method. Whether an actually observed TTVs is due to a Trojan or a non-Trojan planet
is a topic that may require employing other observational techniques, and is outside the
scope of this paper. 

The rest of this paper is outlined as follows. In section 2, we present the results of our stability
analysis and the mapping of the phase space. Section 3 has to do with the calculations of TTVs for different values
of the mass and orbital elements of the transiting and Trojan planets. Section 4 concludes this 
study by reviewing the results and discussing their implications.

\section{Numerical Set Up and Stability Analysis}

We consider a co-planar system consisting of a star, a transiting planet, and a perturbing boy.
The mass of the central star is chosen to be between 0.1 and 1 solar-masses $({M_\odot})$.
The transiting planet is considered to be in a circular orbit with a mass ranging from 0.1 
to 1 Jupiter-mass $(M_{\rm J})$. The orbital period of this object is taken to be between 
3 and 10 days. The perturbing body is a super-Earth $(1 M_\oplus \leq {m_{\rm p}} \leq 10 M_\oplus)$ 
and  its initial orbital eccentricity is chosen from the range of 0 to 0.8 in steps of 0.05. 

For the sake of completeness, we start by studying the stability of the two planets
in a 1:1 MMR. We used the chaos indicator MEGNO 
(Mean Exponential Growth factor of Nearby Orbits, Cincotta and Sim\'o 2000, Go\'zdziewski 2001)
and integrated the three-body system of the star and the transiting and perturbing planets
for different combinations of the mass and mean-anomaly of the two planets. 
Integrations were carried out for more than 100,000 orbits of the transiting body
and were terminated when MEGNO reached a value of 10 or larger. For quasi-periodic orbits,
MEGNO will asymptotically approach the value of 2.0 and it diverges away from 2.0 exponentially
in the case of chaotic orbits. More details on MEGNO and its application to orbital
stability of our systems can be found in the appendix. 

Figures 1 and 2
show samples of the results for different initial configurations of the two planets. 
The central stars in these figures are 1 ${M_\odot}$ and 0.3 ${M_\odot}$, 
respectively. The transiting planet in each system is Jupiter-mass and in a 4-day orbit in figure 1
and 7.5-day orbit in figure 2. The perturbing body has a mass of 1 $M_\oplus$.
Both three-body systems were considered to be co-planar and the initial values of the longitudes 
of ascending nodes and arguments of pericenters of the transiting and perturbing planets were set to zero. For more
studies on the stability of co-orbital planets and their configurations and orbital
elements, we refer the reader to Hadjidemetriou et al (2009, 2011) and Giuppone et al (2010).
 
Integrations were carried out for different values of the semimajor axis and eccentricity
of the perturbing Earth-mass planet. The black curves in each figure show the loci of points for which the
orbital perihelion and aphelion distances are at 0.05 AU from the central star.
As shown in these figures, in general, the orbit of the perturbing planet is stable when it is entirely outside the 
influence zone of the transiting body. As expected, however, within the unstable regions, islands of
stability appear where the perturbing and transiting planets are in mean-motion resonances.
A few of these resonances are shown in figures 1 and 2. 
We refer the reader to Haghighipour and Kirste (2011) for a detailed analysis of the detectability of 
an Earth-mass planet with the TTV method in these resonances. 

We also carried out similar integrations considering
the initial argument of the periastron of the Earth-mass planet to be $60^\circ$. 
A sample of the results, as shown in figure 3,
once again indicate that the orbit of the Earth-mass perturber will be stable in close distances to the central
star. Islands of stability also appear in this case, in particular for the 1:1 MMR, which shows the Earth-mass planet
will be stable for its value of orbital eccentricity up to $\sim 0.4$.

An interesting result shown in figures 1 and 2 is the island of stability at the 1:1 MMR.
As shown here, the orbit of the Earth-mass perturber may be stable for different values of 
its eccentricity. We carried out more detailed analysis of the long-term stability of this
object and integrated our system using the Bulirsch-Stoer integrator of the N-body 
integration package MERCURY (Chambers 1999). 
Figure 4 shows the results for two sample cases where the eccentricity of the perturbing Earth-mass planet is 0.4 and 0.8
(the white crosses in the 1:1 stable region of the top-left panel of figure 1), respectively. 
Note that the angular orbital elements of these objects are similar 
to those of the systems in figures 1 and 2.
As shown in the top panels of this figure, the Earth-mass planet maintained its orbit for 
at least one million years suggesting that the quasi-periodic regions in the locations of 
1:1 MMRs in figures 1 and 2 are true representation of stable planetary orbits.
The bottom panels of this figure show the astrocentric orbits 
of the two planets for one orbit.

We also explored the effect of the orbital eccentricity of the transiting planet on the stability of
the Earth-mass object. For these calculations we considered two cases;
the real system of M star KOI-254 where a $0.505 \, M_{\rm J}$ planet transits a 0.59 solar-mass star
in a 2.45-day orbit with an eccentricity of 0.1 (Johnson et al 2011), and a system consisting of a 
solar-mass star with a transiting Jupiter-mass planet in a 3-day circular orbit. 
Figures 5 and 6 show the MEGNO maps of an Earth-mass planets in these systems. As seen from figure 5, 
the slight eccentricity of the transiting planet KOI-254 b has caused almost the entire region of its 
associated 1:1 MMR to become unstable. However, figure 6 shows that in the system of this figure,
an Earth-mass planet can have stable orbits in both ${\rm L}_4$  and ${\rm L}_5$ Lagrangian points 
when the eccentricities of both planets are 0.1.

\section{Calculation of TTV}

To calculate the variations in the transit timing of the giant planet, we first performed similar
N-body integrations as those explained above, once with and once without the perturbing Earth-mass body. Integrations
were carried out for a time span of 30 years to show both the short-term and long-term effects
(we note that for the purpose of detecting TTVs with the {\it Kepler}, a time span of 3 years
will suffice). For each set of simulations, the transit timing 
variations were calculated by obtaining the difference between the time of mid-transit in the systems
with and without the perturbing body. We assumed that at $t=0$, 
the centers of the star and the transiting planet were on the $x$-axis, and calculated 
the time of transit by interpolating between the
times before and after the center of transiting planet crossed the center of the
star. 

Figure 7 shows a sample of the results. The system in this figure consists of a solar-mass
star, a Jupiter-mass transiting planet in a 3-day orbit, and an Earth-mass perturber in the 
${\rm L}_4$ Lagrangian point. The four panels correspond to the initial orbital eccentricity
of the Trojan planet from 0 to 0.15. The initial orbital eccentricity of the transiting planet
was set to zero. As shown here, the maximum values of TTV vary from $\sim 16$ s  
to $\sim 72$ s by increasing the eccentricity of the Earth-mass planet.
A comparison between the amplitudes of TTVs in figure 7 and the transit timing variations of the planetary candidates
identified by the {\it Kepler} space telescope in Q0-2 indicates that these TTVs fall in the lowest range of the detected 
transit timings [See figure 1 of Ford et al. (2011); The smallest detected TTV in this figure is
slightly larger than 12 seconds]. 

Figure 7 also shows that as expected, an increase in
the orbital eccentricity of the Trojan planet results in an increase in the amplitude of TTVs. 
This motivated us to examine the change in the magnitude of TTVs for different values of
the initial orbital eccentricity of the perturbing planet. As shown in figure 3, 
the orbit of the Trojan planet becomes unstable for large values of its eccentricity. We, therefore, chose
a system similar to that of the top-left panel of figure 1, and calculated TTVs for different values 
of the orbital eccentricity of the Trojan perturber. Figure 8 shows the results for a time span 
close to 30 years. 
As shown here, the amplitudes of TTVs are now within the range of approximately 160 s to 360 s. 
This range is within the lower limit of the TTVs of the {\it Kepler's} multiple planet candidates 
($\sim 2$ minutes, see figure 1 of Ford et al. 2011) pointing to the large prospect for the detection of 
these objects within the extended lifetime of the {\it Kepler}.

We performed similar calculations for different values of the mass of the perturbing body ($1-10$ Earth-masses), 
its orbital eccentricity ($0 - 0.15$), and the mass and orbital period of the transiting planet.
Figure 9 shows the maximum values
of the TTVs of a Jupiter-mass planet in a 3-day orbit for four different values of the mass
of the Trojan perturber. The central star is solar-mass and the transiting planet was initially 
in a circular orbit. As expected, the magnitude of TTVs increase, reaching to several minutes, 
by increasing the mass and eccentricity of the Trojan body. These values are  
well within the range of the {\it Kepler's} detected TTVs ($\sim 8$ minutes, Ford et al. 2011)
implying reasonable probability 
of detection for super-Earth Trojans with slight orbital eccentricities. Figure 10 shows similar
calculations when the mass of the transiting planet is 0.1 $M_{\rm J}$ and 0.5 $M_{\rm J}$. In these systems,
even circular super-Earths induce TTVs that are comparable with the TTVs of some of the {\it Kepler}'s
planets such as Kepler-9 b\&c ($\sim$ 25 minutes to a slightly more than an hours, Holman et al 2010). 

The amplitudes of TTVs are significantly increased when the transiting planet  
is at larger orbits. As shown in figure 11, the maximum values of
TTVs vary between approximately 0.6 and 2.3 hours when a 0.1 $M_{\rm J}$ transiting planet in
orbits with periods between 3 and 10 days is perturbed by a 6 Earth-mass super-Earth Trojan.
Such large TTVs are entirely within the bulk of the TTVs of the {\it Kepler}'s 
planetary candidates ($10 - 40$ minutes, Ford et al. 2011).
Figure 12 shows this trend in terms of the mass of the Trojan planet and for different values
of the orbital period of the transiting body where the amplitude of TTVs raise up to over 4 hours. 
Results indicate that in general, super-Earth
Trojans with small orbital eccentricities have a reasonably high probability of detection 
in systems where the transiting planets are slightly smaller than Jupiter and in moderately
short-period orbits.

\section{Concluding Remarks}
We have studied the detectability of Earth-mass Trojans by comparing the variations
they induce in the transit timing of their Jovian-type host planets with the detected
transit timing variations of {\it Kepler's} planetary candidates. We analyzed the stability of
these systems for different values of the mass and orbital elements of the transiting and perturbing
bodies, and calculated their associated TTVs for the ranges of their orbital elements that
correspond to their long-term stability. Results indicated that while in general, the
amplitude of TTVs fall within the detectable range of timing precision obtained from the {\it Kepler}'s
long-cadence data, the prospects of the detection of Trojan planets are higher for super-Earth Trojans 
in slightly eccentric orbits around transiting Jovian planets with masses smaller than Jupiter. 
Our study also present a pathway to break the strong degeneracy associated with the RV-detection 
of Trojan planets around $L_4$ and $L_5$ (Giuppone et al 2012) by combining the
information obtained from the study of the transits of their systems and their TTVs.

The purpose of our study was to identify regions of the parameter space for which
the amplitude of the TTVs of a transiting planet due to the perturbation of its Trojan body
would fall within the range of the photometric sensitivity of the {\it Kepler}.
However, whether a detected TTV is actually due to a Trojan planet cannot be deduced from our results.
While as shown by Nesvorn\'y and Morbidelli (2008), Nesvorn\'y (2009), Nesvorn\'y and Beauge (2010),
and Nesvorn\'y et al (2012), it would be possible in non-resonant systems to determine the mass 
and orbital elements of the perturbing body by analyzing TTVs alone, to extract the 
nature of a perturbing Trojan planet from detected TTVs, one has to either 
compare the detected TTVs with those in large TTV  catalogs (e.g., Haghighipour and Kirste 2011)
or complement the detection by using other planet-detection techniques, in particular 
the radial velocity method (Laughlin and Chambers 2002, Giuppone et al 2012).

We based our study on the assumption that similar to the giant planets 
of our solar system, extrasolar giant planets may also have Trojans objects and satellites. 
These objects may be as large as terrestrial planets, in which case their perturbations on
the orbital motion of their host planet may become detectable. How such Trojan planets
form is a matter of on-going research. In general, one can think of three scenarios; 
in-situ formation around ${\rm L}_4$ and ${\rm L}_5$ Lagrangian points of a short-period
giant planet, in-situ formation around the host giant planet followed by the inward migration of
the planet-Trojan system, and formation in the inner part of the protoplanetary disk followed 
by capture in a 1:1 MMR during the migration of a giant planet. As these scenarios are general,
they present variety of possibilities for incorporating formation, migration, and capture
mechanisms into a single model. We refer the reader to studies by Laughlin and Chambers (2002),
Chiang and Lithwick (2005), Thommes (2005), Cresswell and Nelson (2006), Beaug\'e et al (2007), 
Morbidelli et al (2008), Hadjidemetriou and Voyatzi (2011), and Giuppone et al (2012)
for more details on the models of Trojan planet formation, and the current state of research on this topic.

\begin{acknowledgements}

NH acknowledges support from NASA EXOB grant NNX09AN05G, the NASA Astrobiology Institute under 
Cooperative Agreement NNA09DA77A at the Institute for Astronomy (IfA), University of Hawaii,
and Alexander von Humboldt Foundation. NH is also thankful to the Computational Physics group at the Institute 
for Astronomy and Astrophysics, University of T\"ubingen for their kind hospitality during the course
of this project. SC acknowledges support from the IfA NSF-funded REU program. 
TCH acknowledges support from the Korea Astronomy and Space Science
Institute (KASI) grant 2012-1-410-02 and the Korea Research Council for
Fundamental Science and Technology (KRCF) through the Young Research
Scientist Fellowship Program. The MEGNO computations were carried out at the
SFI/HEA Irish Center for High-End Computing (ICHEC) Center and the
PLUTO computing cluster at KASI. TCH would also like to thank the 
Institute for Astronomy and the NASA Astrobiology Institute at the University of Hawaii-Manoa 
for their hospitality during the course of this project.

\end{acknowledgements}

\clearpage

\begin{figure}
\centering{
\includegraphics[scale=0.48]{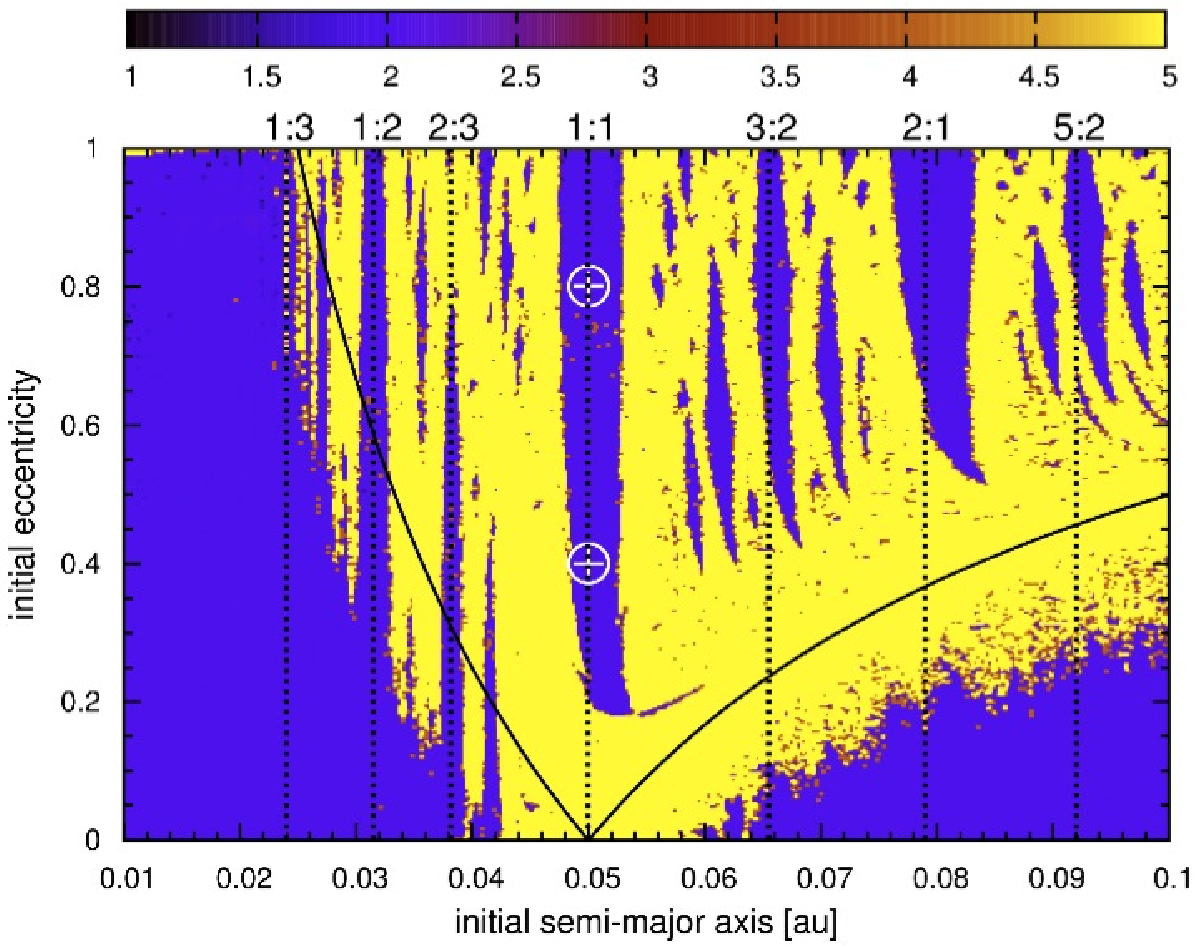}
\includegraphics[scale=0.48]{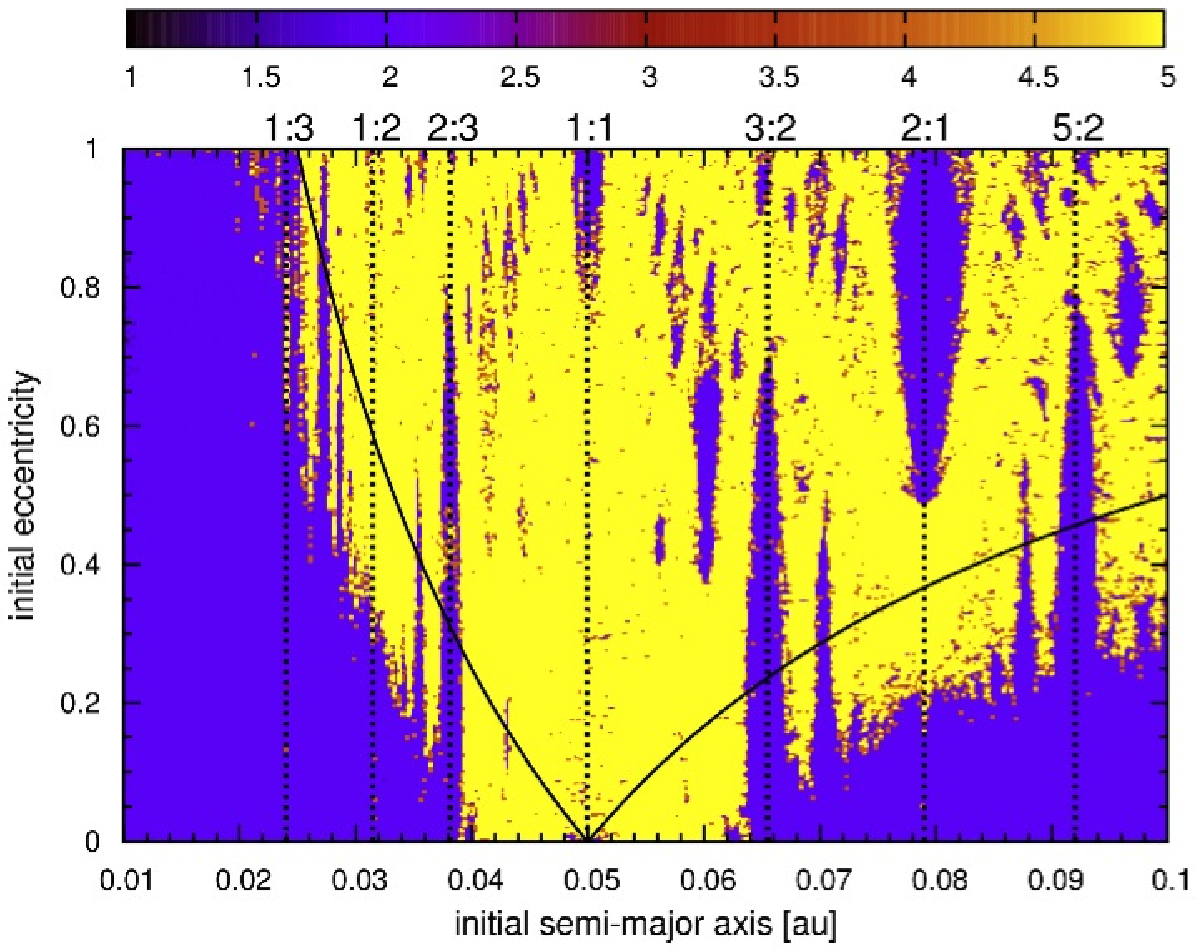}
\vskip 10pt
\includegraphics[scale=0.48]{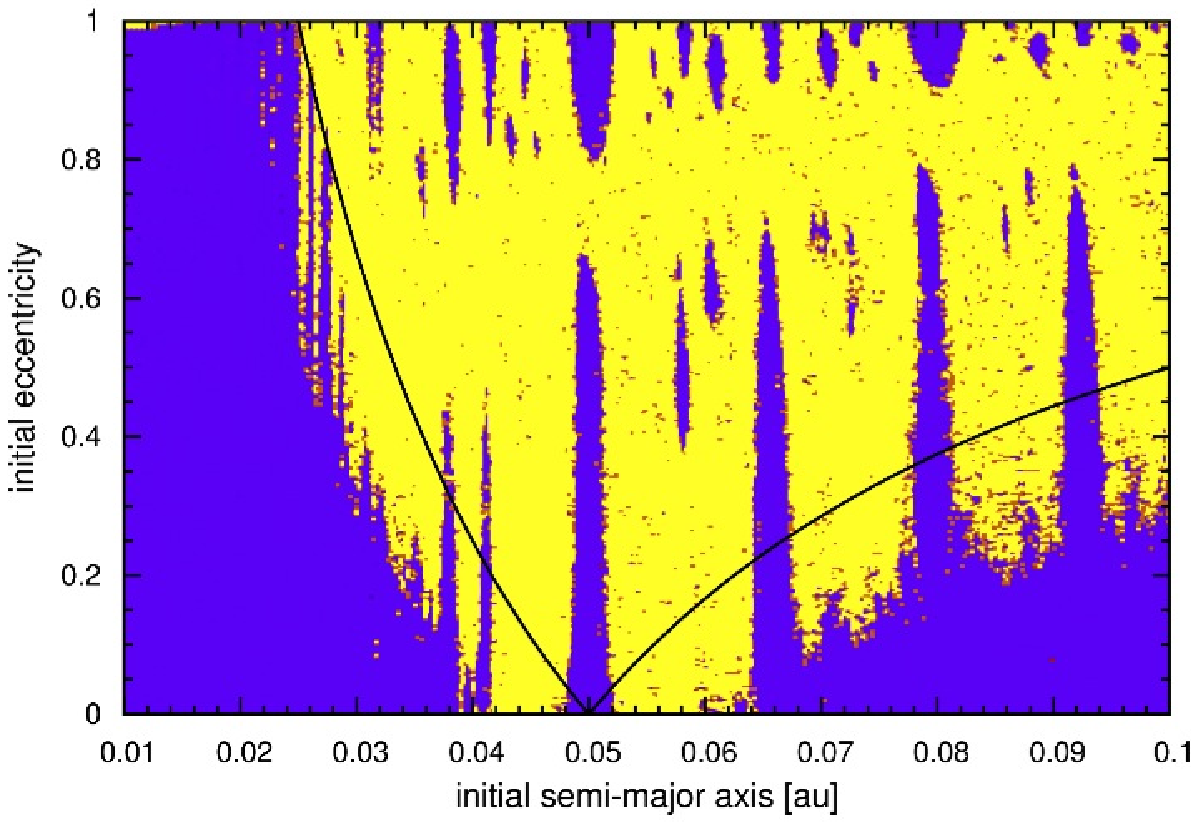}
\includegraphics[scale=0.48]{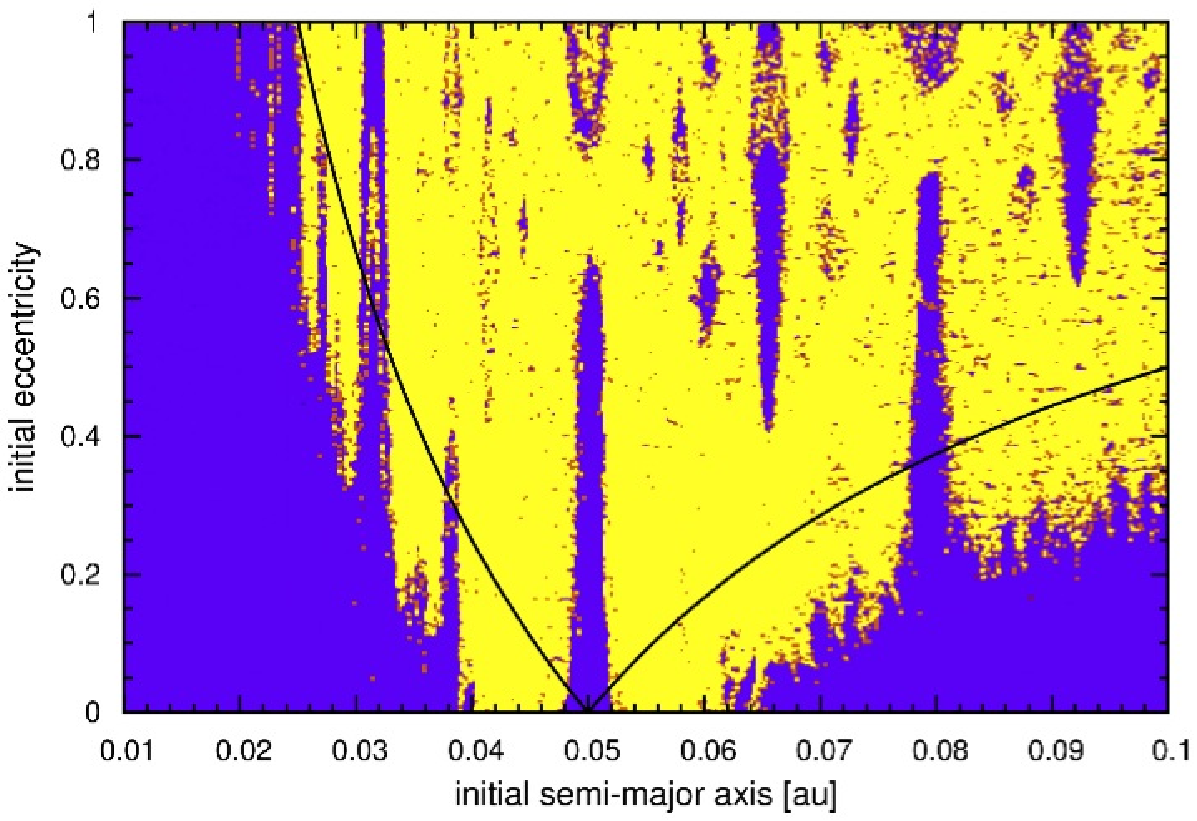}
\vskip 10pt
\includegraphics[scale=0.48]{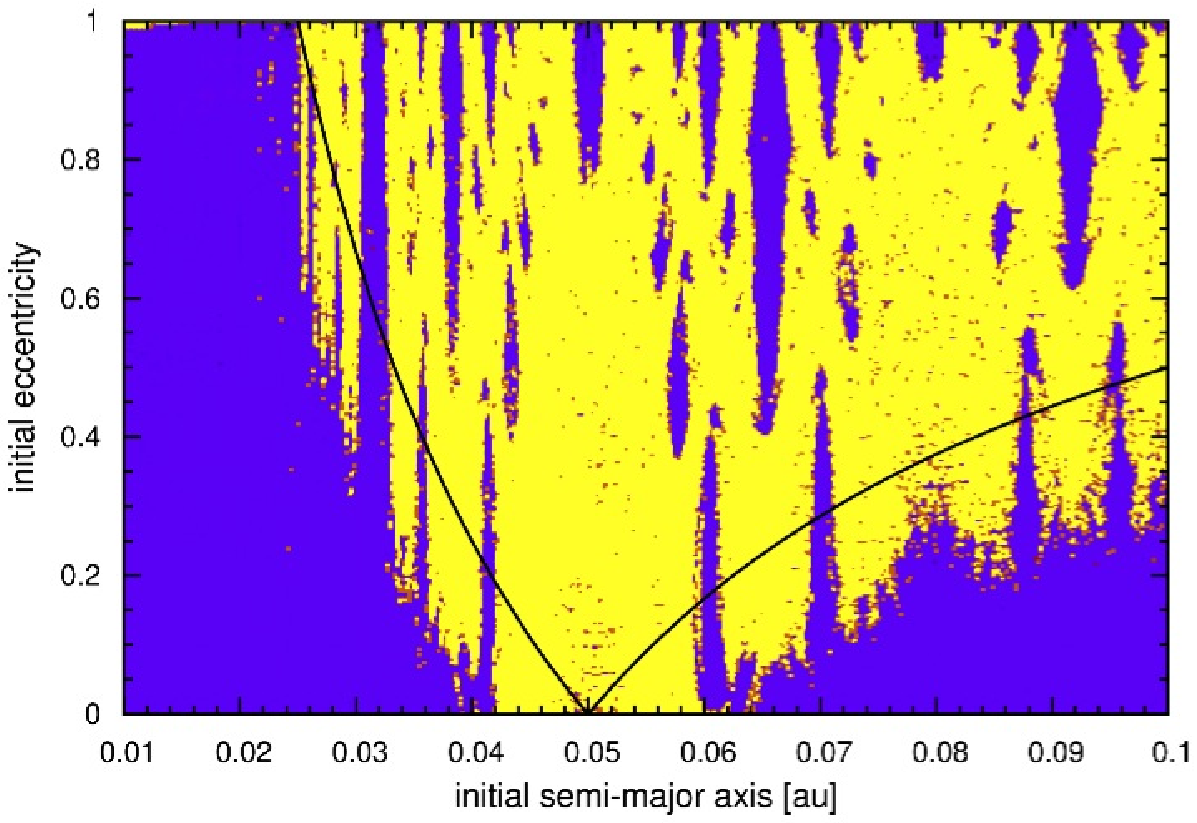}
\includegraphics[scale=0.48]{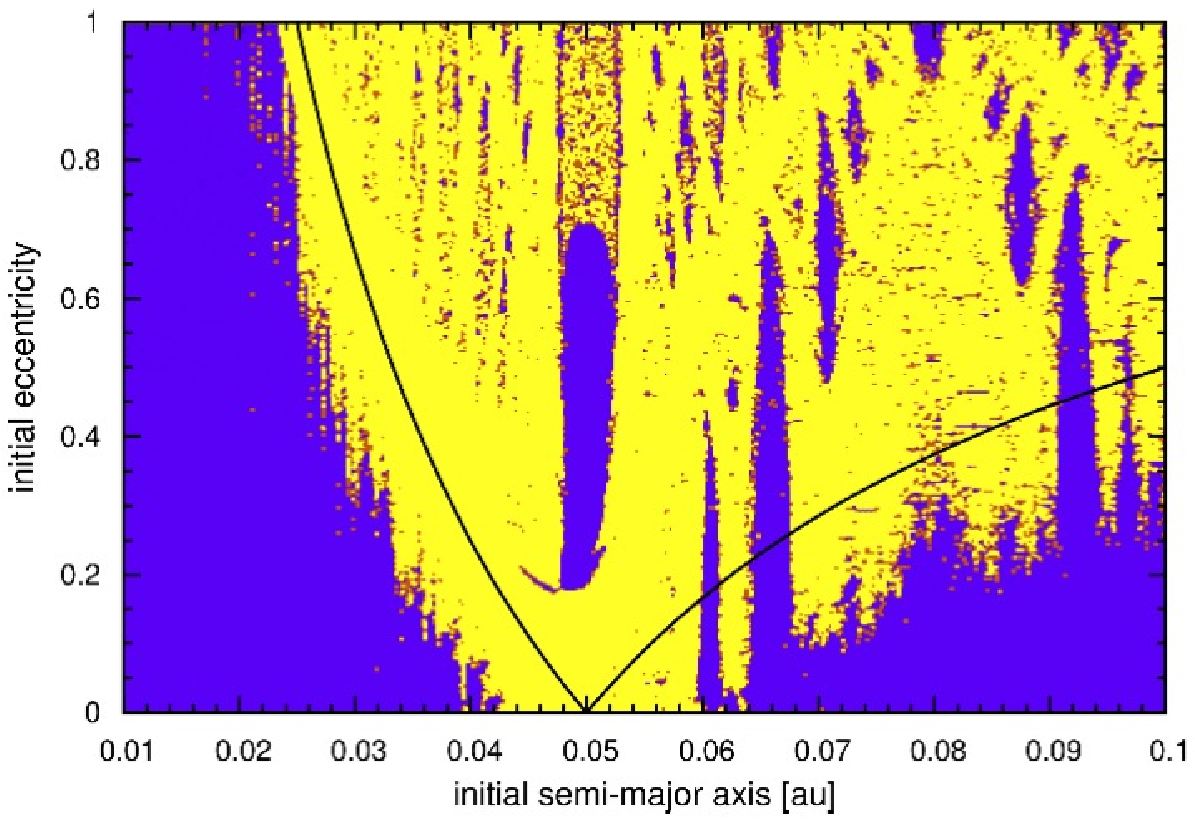}}
\caption{MEGNO dynamical maps for an Earth-mass perturbing planet in a system consisting of a solar-mass
star and a Jupiter-mass transiting body. The vertical and horizontal axes show the initial values
of the orbital eccentricity and semimajor axis of the Earth-mass planet. The transiting planet 
is in a 4-day orbit and was initially started with no eccentricity. The black curves show the loci
of points whose perihelion/aphelion distances are 0.05 AU from the central star.
The color-coding shows the value of MEGNO from 1 corresponding to a regular orbit to 5 indicating
chaotic behavior. Each panel corresponds to a different combination of the initial mean-anomaly
of the transiting and perturbing planets. In the panels on the left (right), the initial mean-anomaly of
the Earth-mass perturber was set to zero ($180^\circ$) and the mean-anomaly of the transiting planet 
was 0 (top), $90^\circ$ (middle), and $180^\circ$ (bottom), respectively. Several islands of stability 
corresponding to mean-motion resonances between the two planets are also shown.}
\end{figure}

\clearpage

\begin{figure}
\centering{
\includegraphics[scale=0.48]{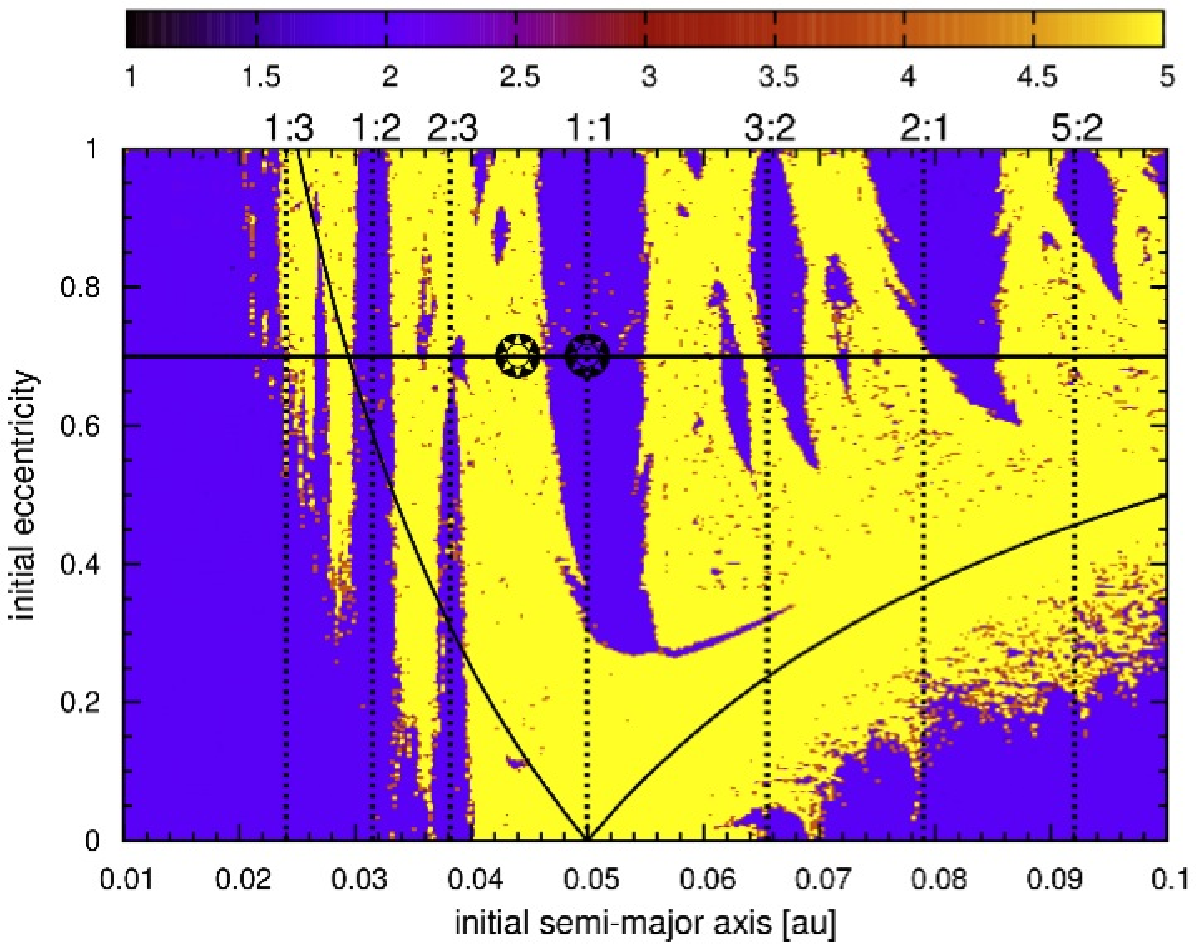}
\includegraphics[scale=0.48]{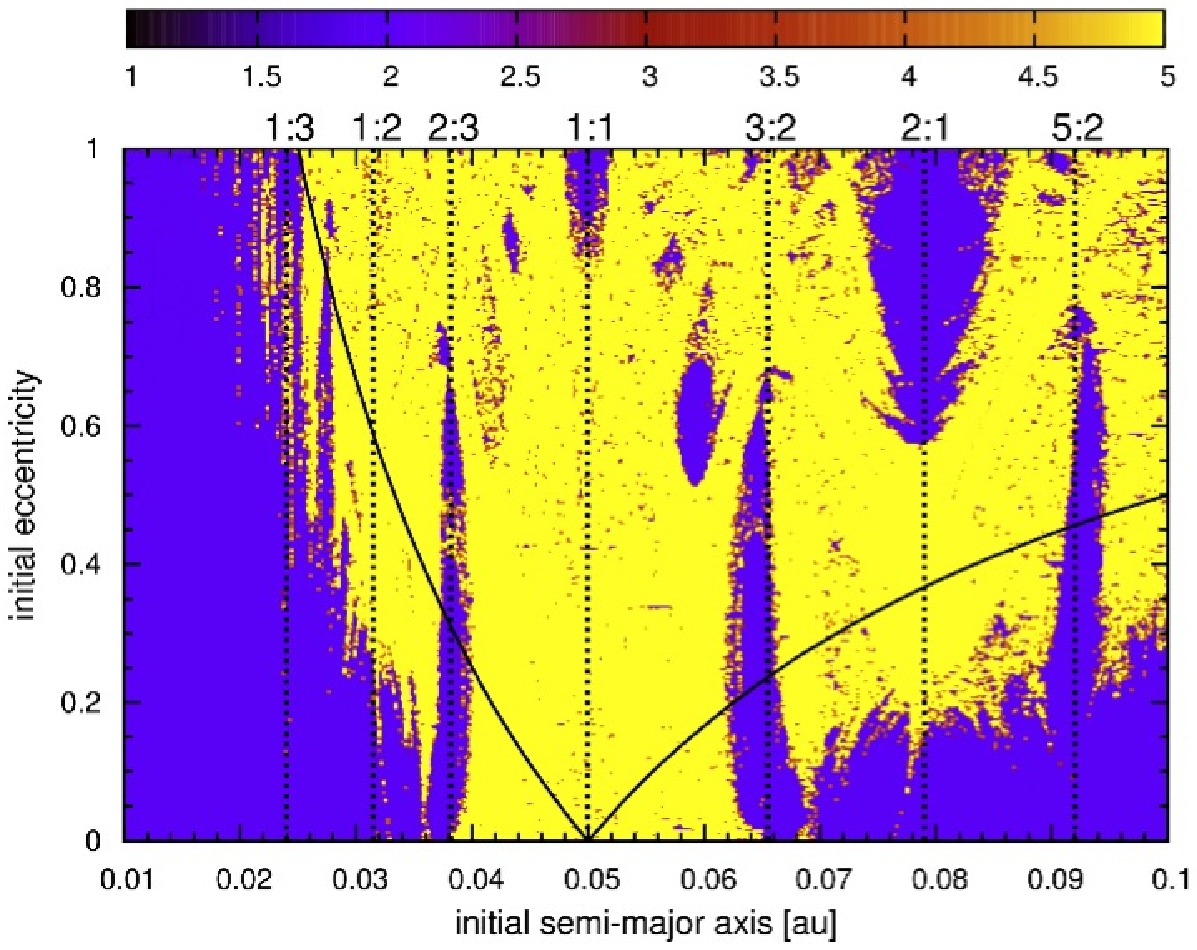}
\vskip 10pt
\includegraphics[scale=0.48]{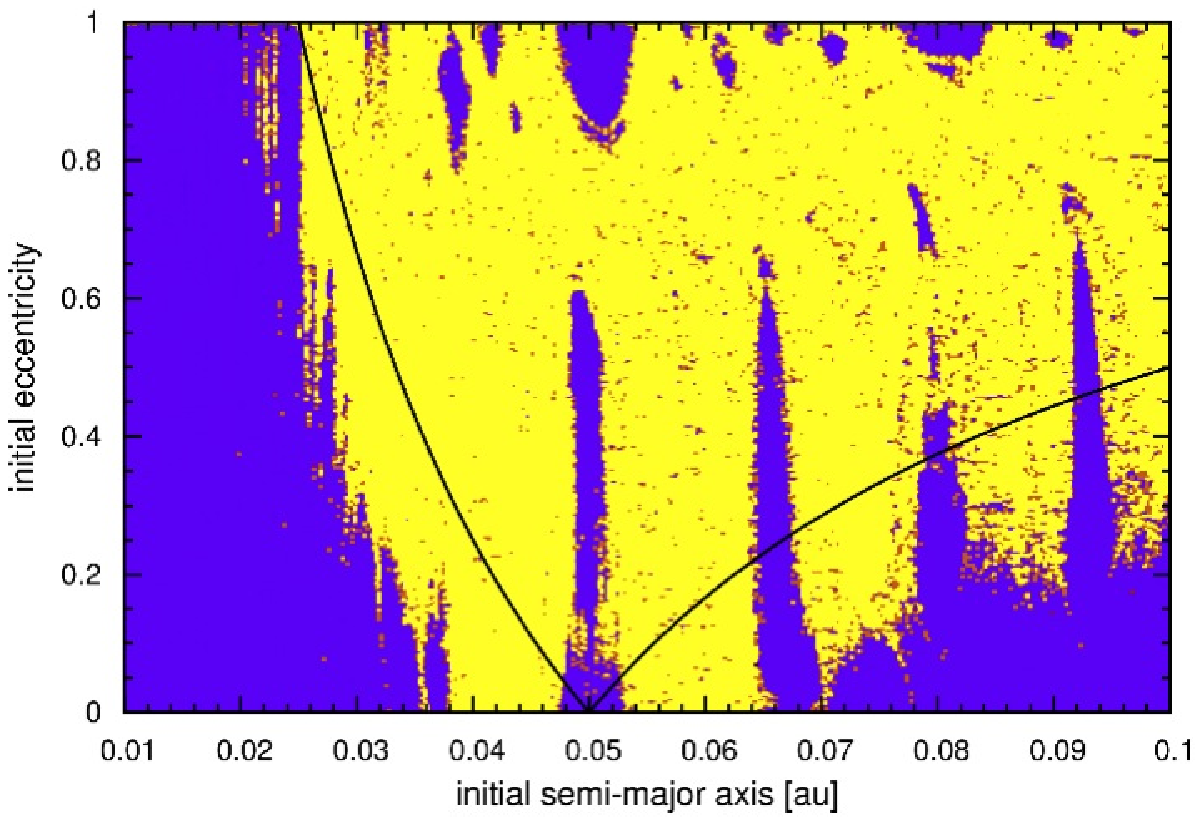}
\includegraphics[scale=0.48]{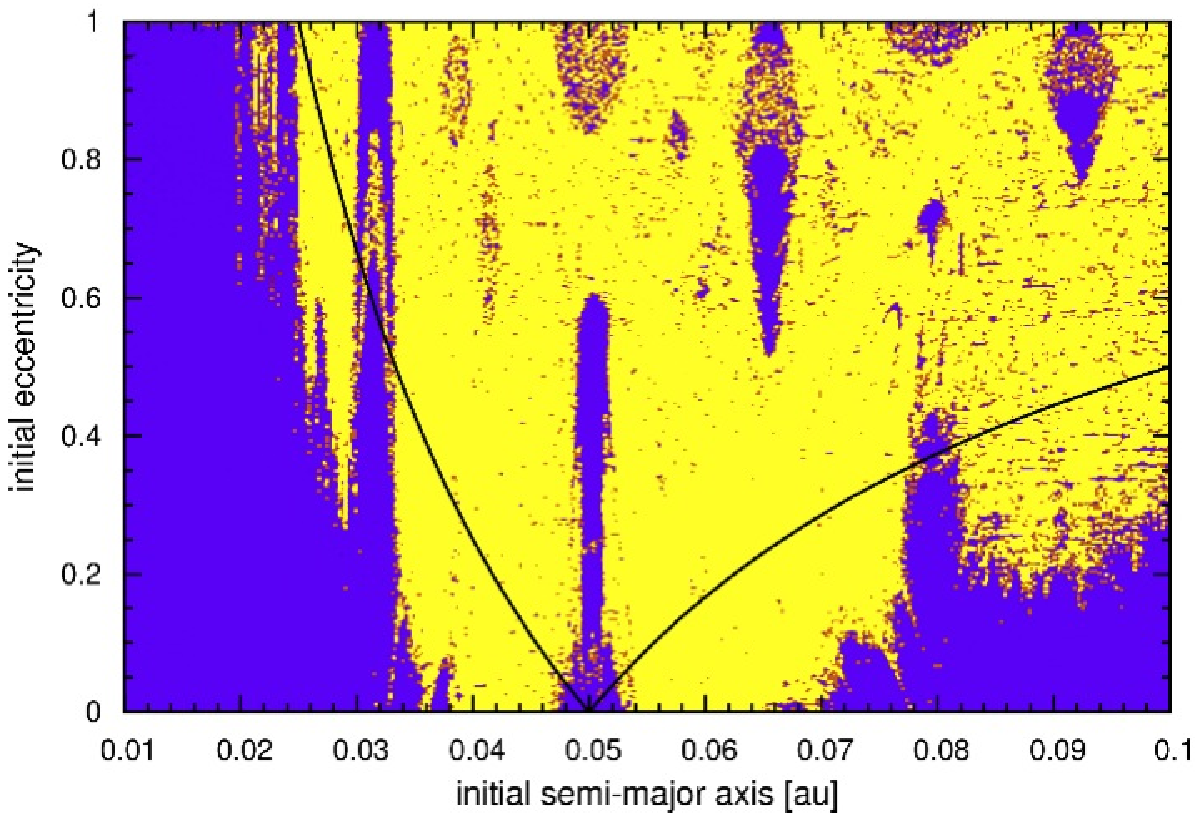}
\vskip 10pt
\includegraphics[scale=0.48]{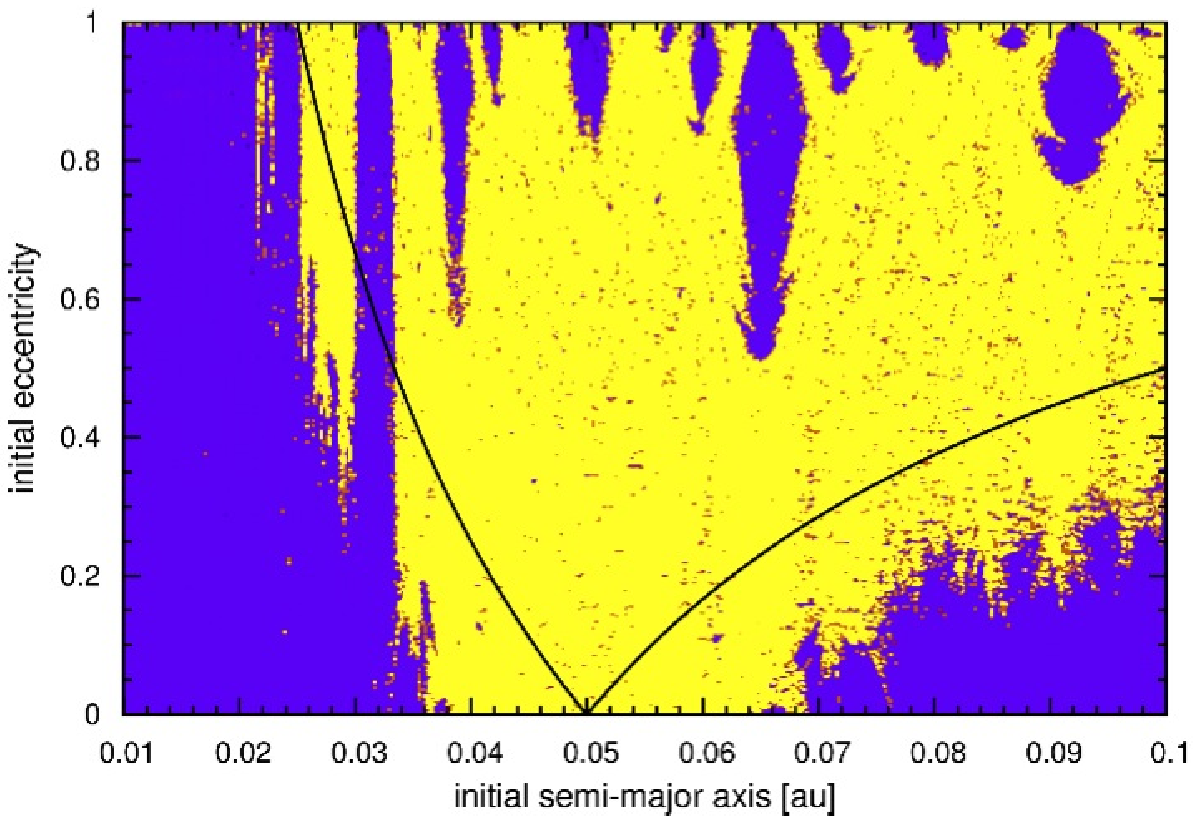}
\includegraphics[scale=0.48]{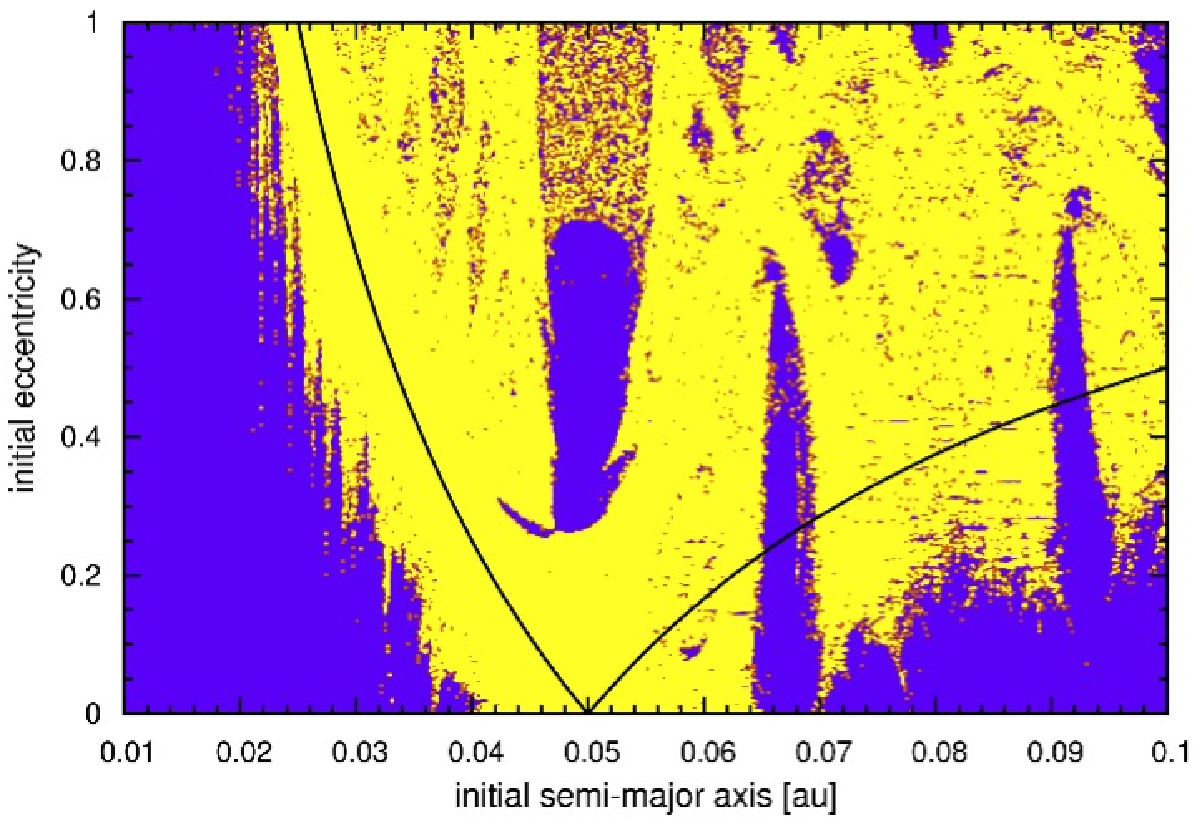}}
\caption{MEGNO dynamical maps similar to Figure 1 for a 0.3 solar-mass star. 
The giant planet is in a 7.5-day orbit.}
\end{figure}

\clearpage

\begin{figure}
\vskip 1in
\centering{
\includegraphics[scale=1]{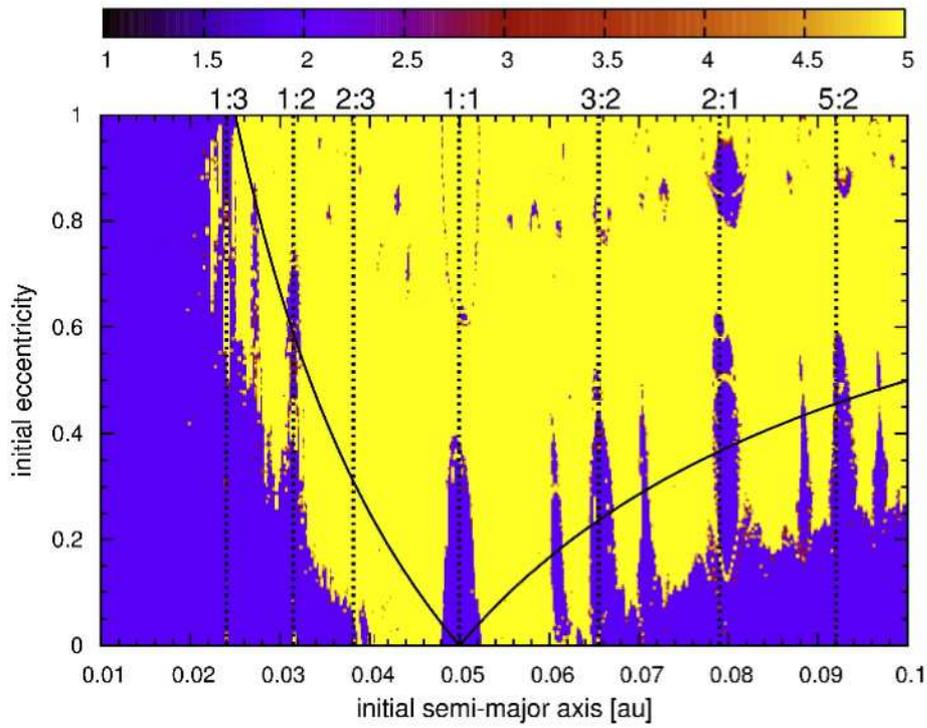}}
\caption{MEGNO dynamical map similar to top-left panel of figure 1 for an Earth-mass planet with an initial
argument of pericenter of $60^\circ$. As shown here, the Trojan planet has a stable orbit
for the initial values of its orbital eccentricity small than $\sim 0.35$.}
\end{figure}

\clearpage

\begin{figure}
\centering{
\includegraphics[scale=0.47]{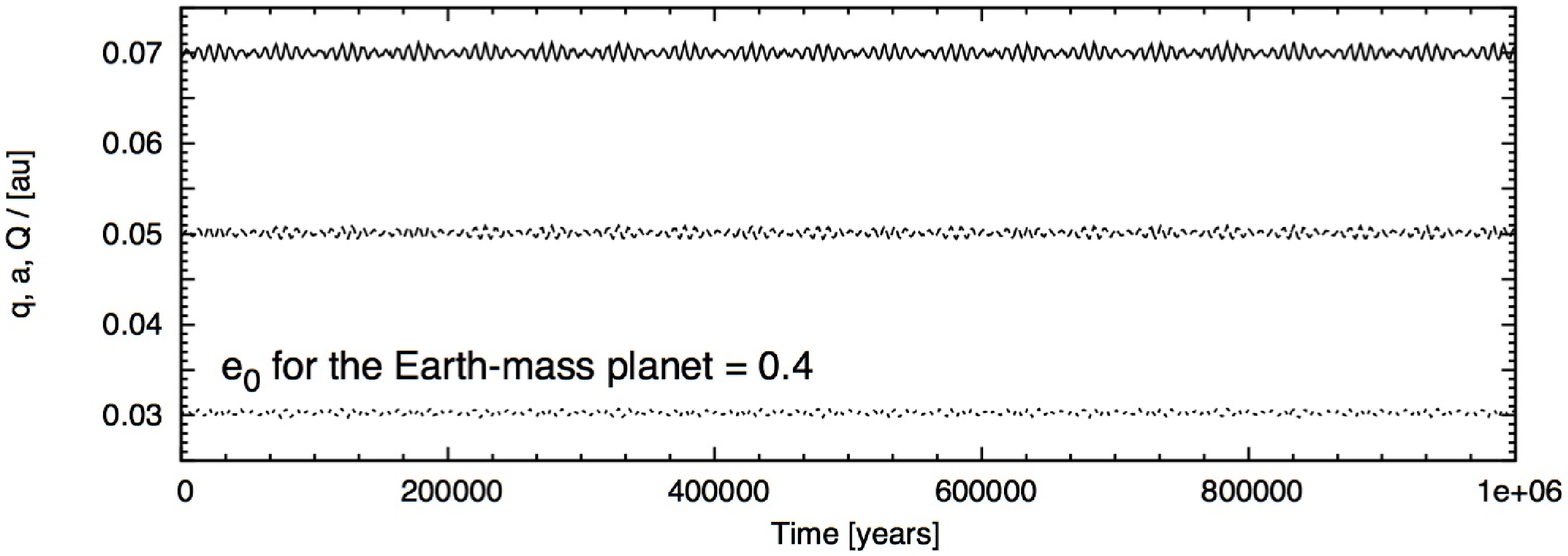}
\includegraphics[scale=0.47]{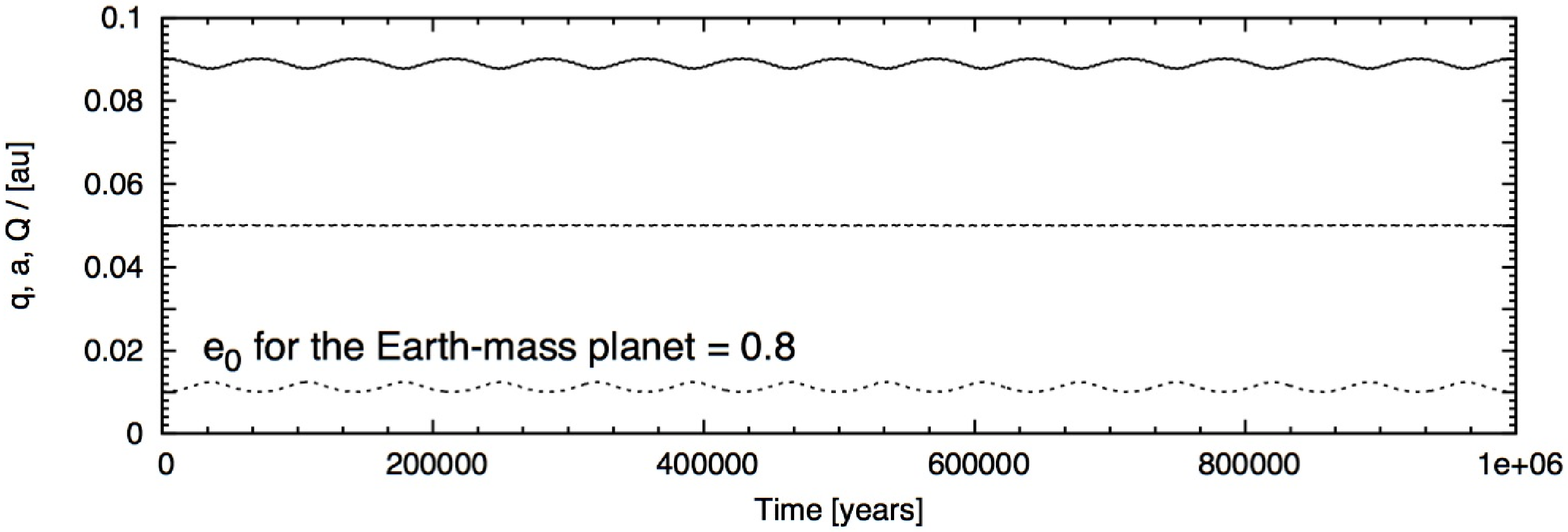}
\vskip 15pt
\includegraphics[scale=0.31]{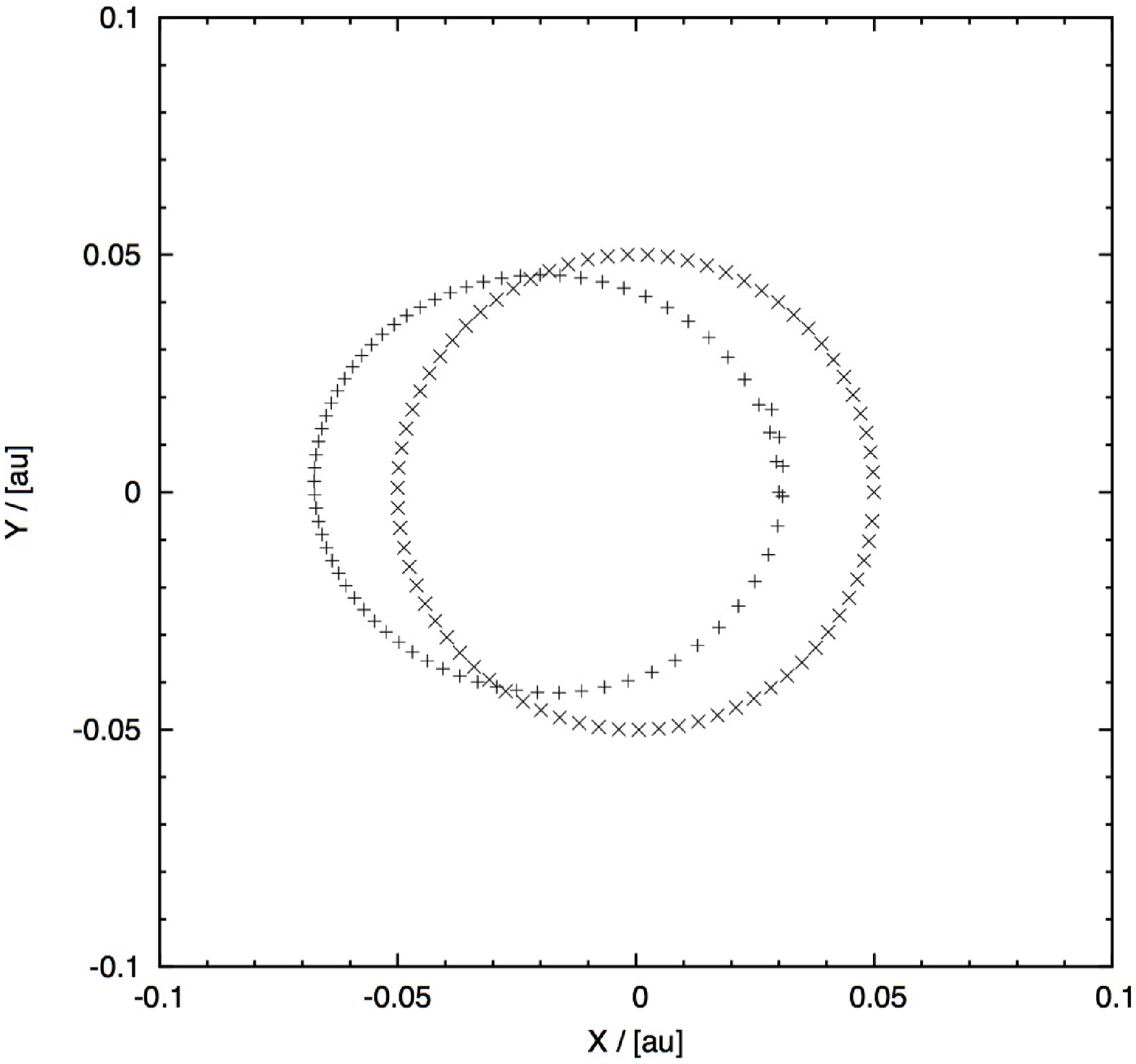}
\includegraphics[scale=0.31]{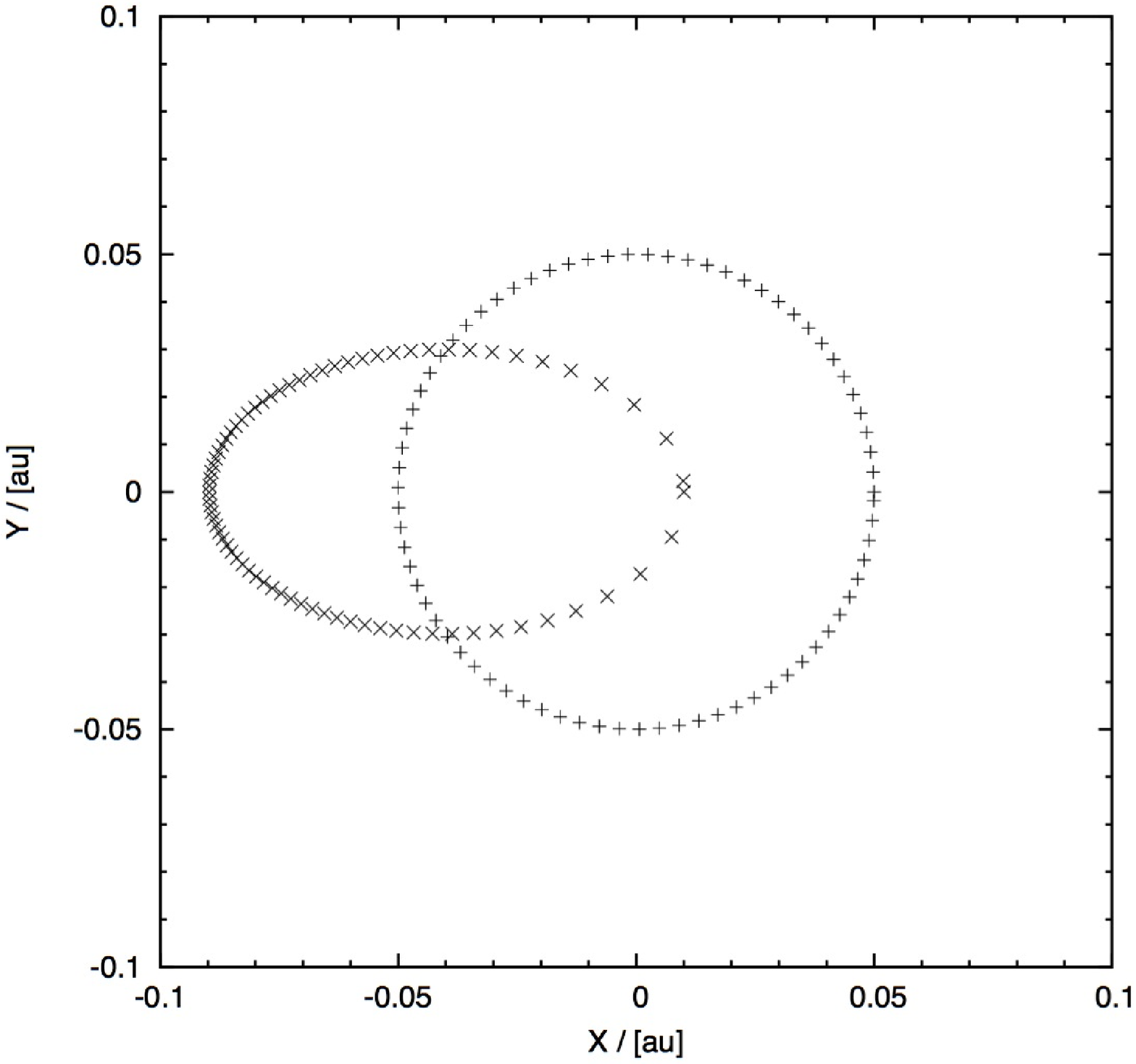}}
\caption{Top: Graph of the semimajor axis $(a)$, perihelion $(q)$, and aphelion $(Q)$ distances
of an Earth-mass perturber in the system of figure 1. The initial eccentricity of the Earth-mass planet
was 0.4 corresponding to the lower white cross in the 1:1 MMR region of the top-left panel of figure 1.
Middle: Similar graph as in the top panel where the initial eccentricity of the Earth-mass planet was 0.8.
The orbit of this planet corresponds to the upper white cross in the 1:1 MMR region of the top-left panel of figure 1.
Bottom: The astrocentric orbits of the transiting giant planet and the Trojan perturber showing only one orbit.
The bottom-left panel corresponds to the system of the top graph, and the bottom-right panel is for the system of the
middle graph.}
\end{figure}

\clearpage

\begin{figure}
\centering{
\includegraphics[scale=0.65]{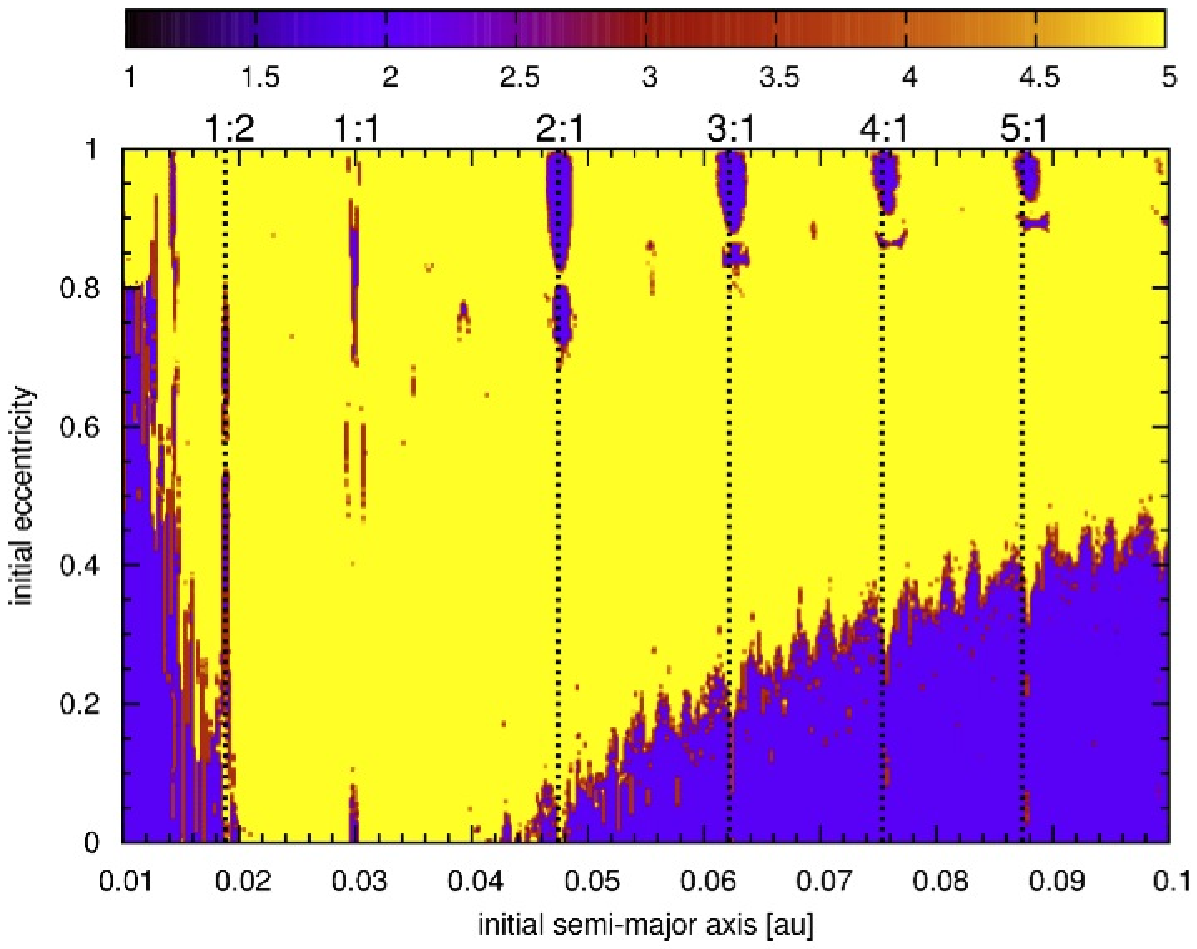}
\vskip 5pt
\includegraphics[scale=0.65]{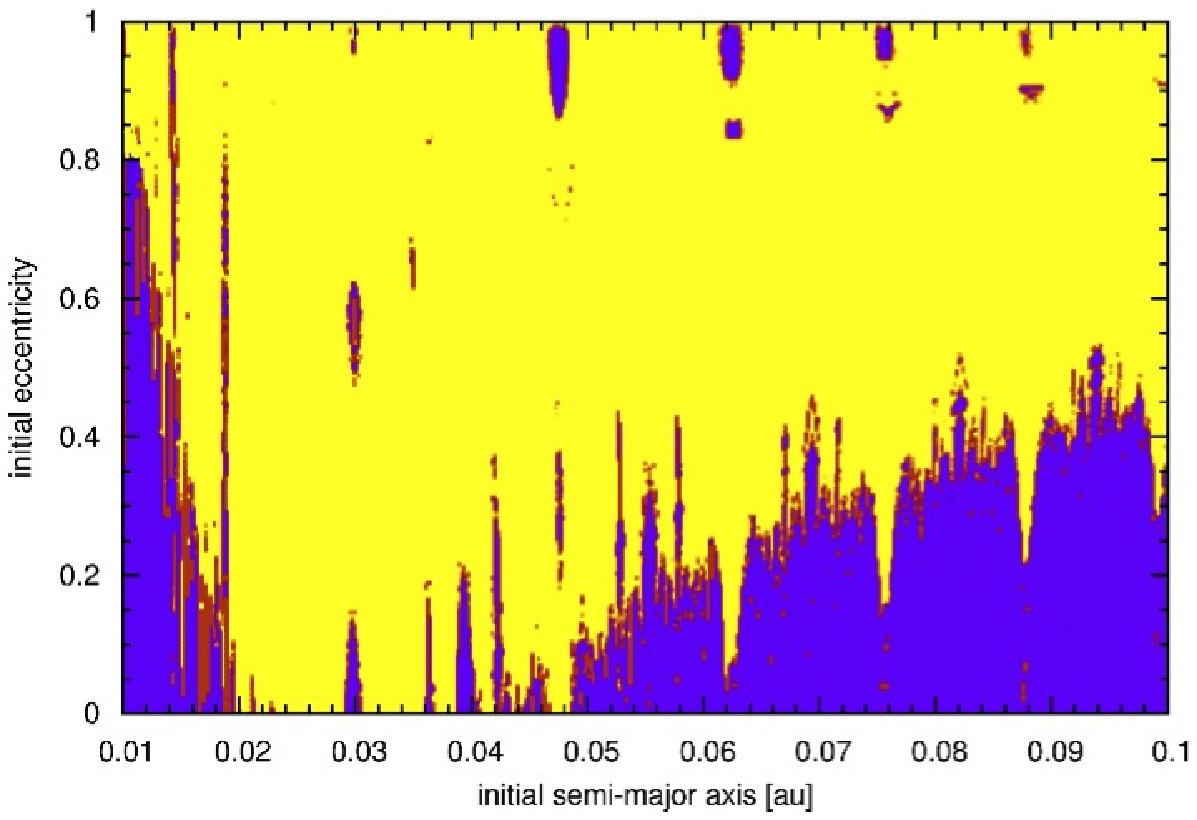}
\vskip 5pt
\includegraphics[scale=0.65]{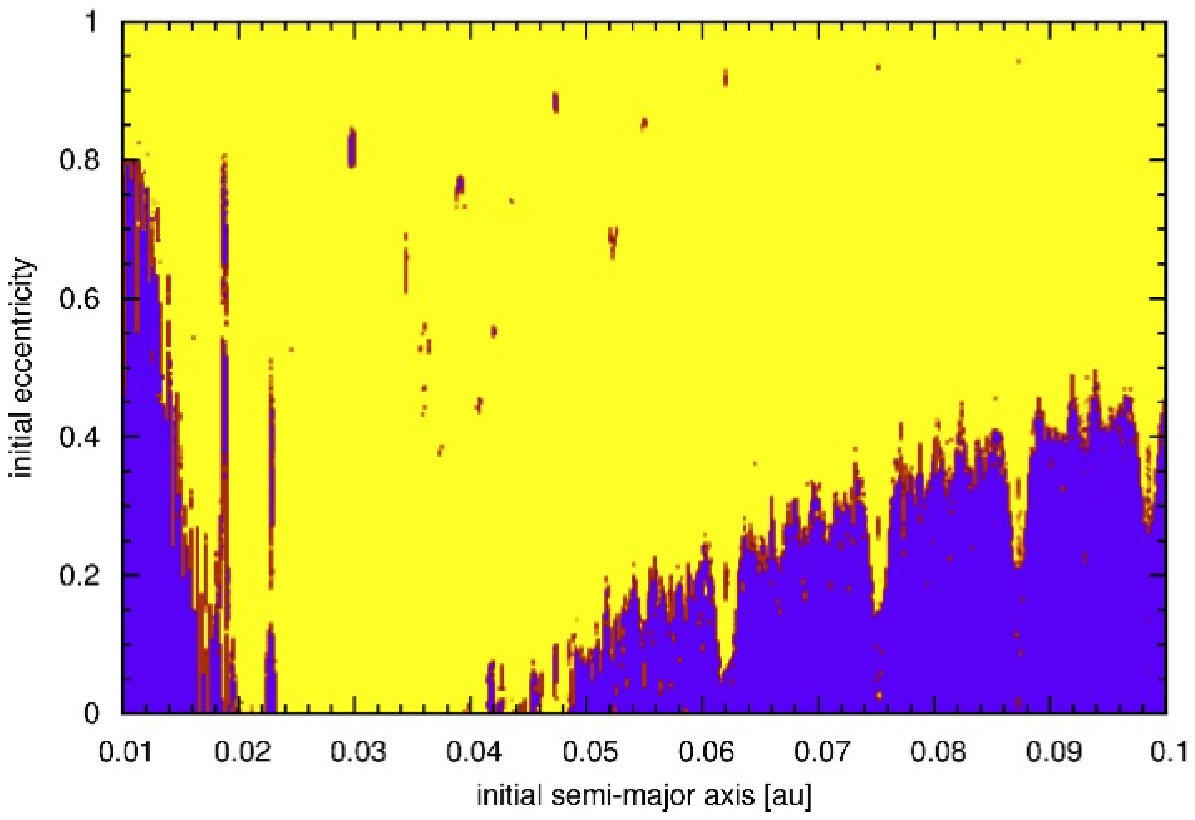}}
\caption{MEGNO dynamical maps for an Earth-mass perturber in the transiting system of KOI-254. The
initial orbital elements of the transiting planet, KOI-254 b (0.505 Jupiter-mass), have been taken from 
Johnson et al (2012). The Earth-mass perturber was initially in a circular orbit with a radius 
equal to that of the transiting planet. Its angular orbital elements were initially set to zero.
The initial mean anomaly of the transiting planet was set to $90^\circ$ (top), $180^\circ$ (middle), 
and $270^\circ$ (bottom). In the top panel, the 2:1, 3:1, 4:1, and 5:1 MMRs have widths of 0.0020,
0.0022, 0.00215, and 0.00158 AU, respectively.}
\end{figure}

\clearpage

\begin{figure}
\centering{
\includegraphics[scale=0.48]{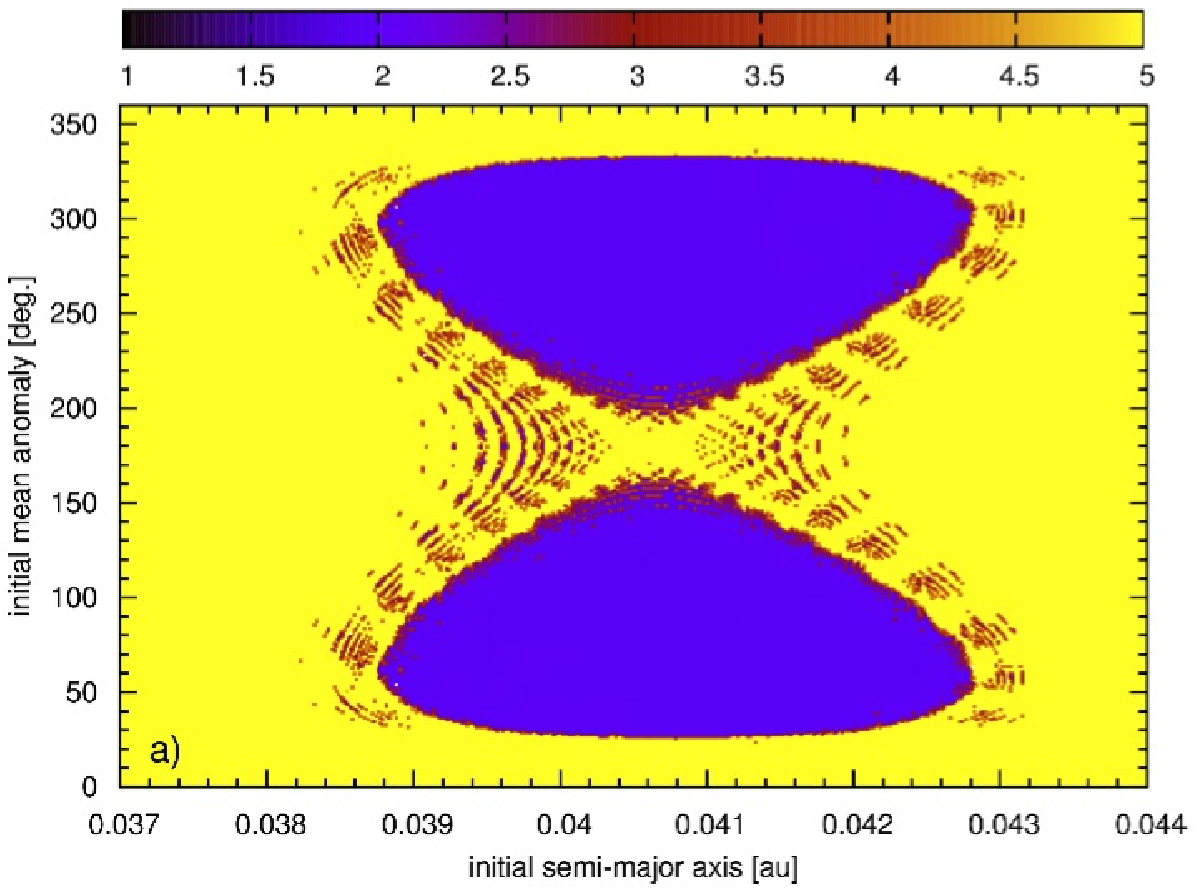}
\includegraphics[scale=0.48]{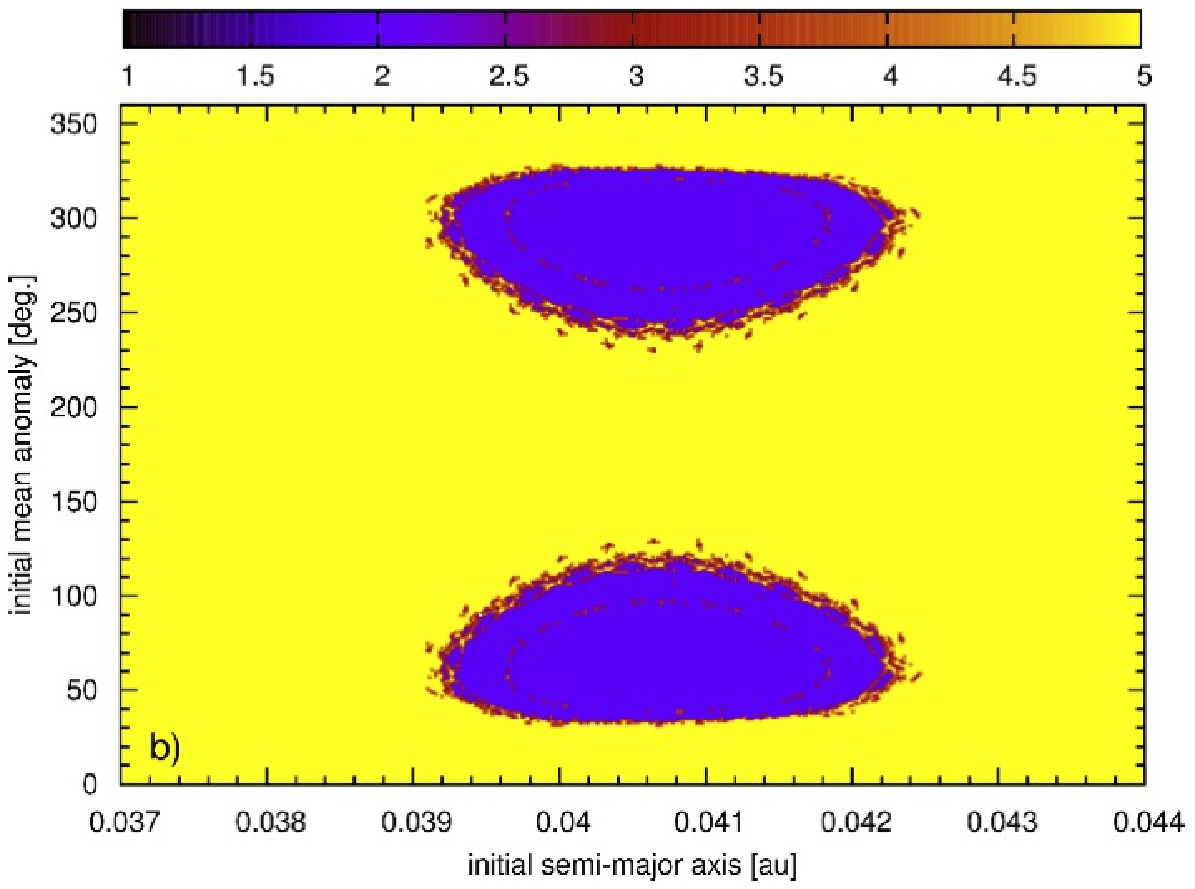}
\vskip 5pt
\includegraphics[scale=0.49]{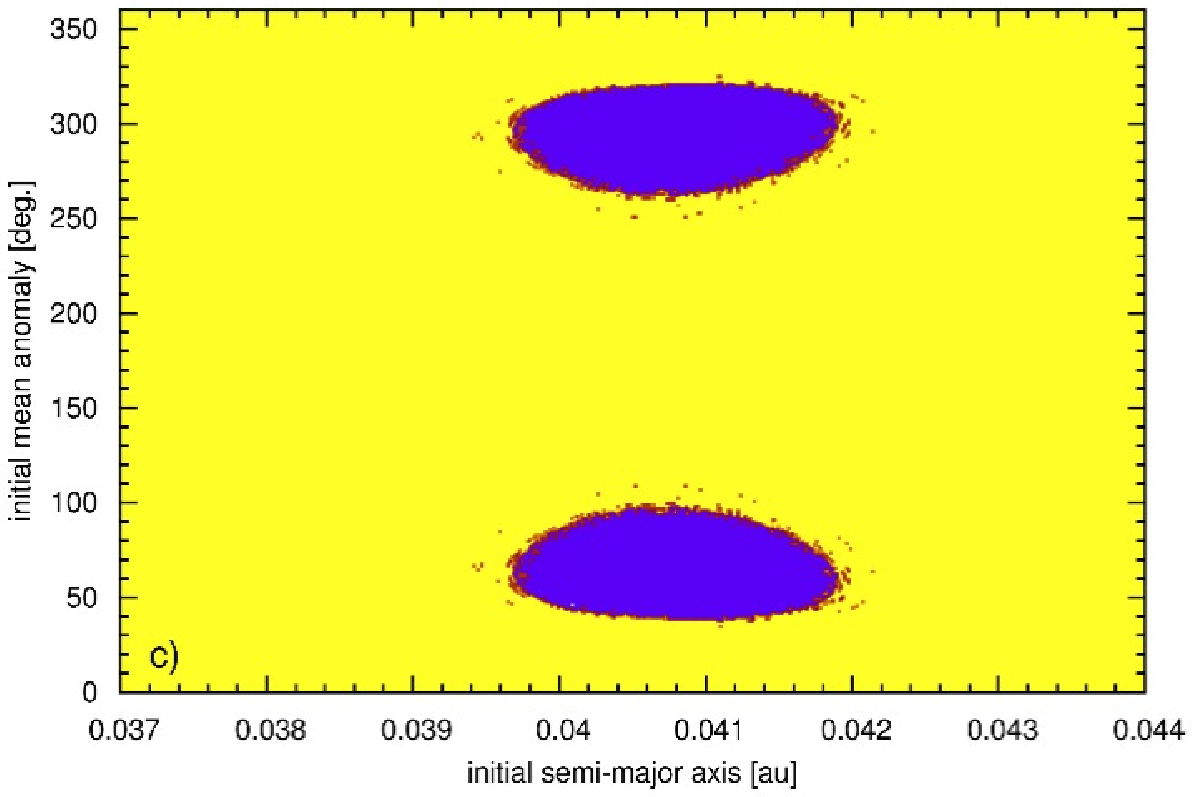}
\includegraphics[scale=0.49]{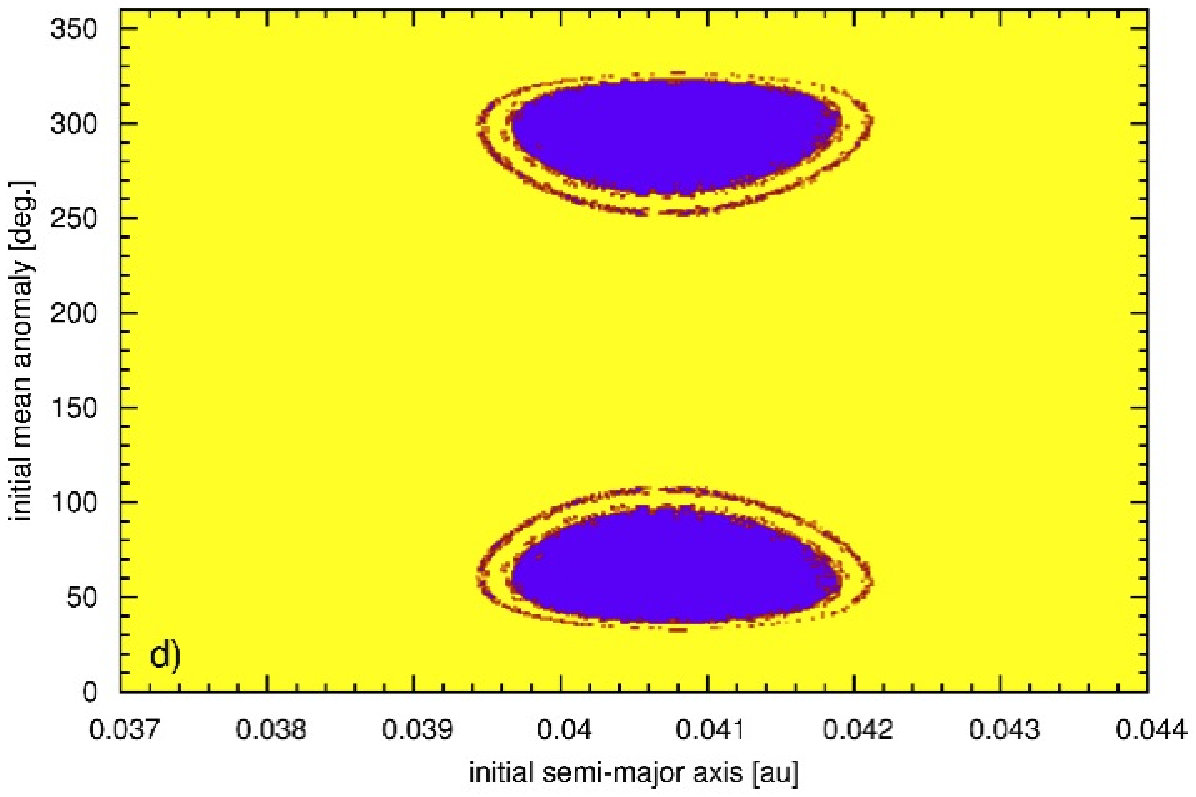}}
\caption{MEGNO dynamical maps for an Earth-mass Trojan in a system consisting of a Jupiter-mass transiting
planet in a 3-day orbit and a solar-mass star. The initial eccentricities of the (transiting , Trojan)
planets were set to (0 , 0) for the top-left panel, (0 , 0.1) for the top-right panel, (0.1 , 0) for
the bottom-left panel, and (0.1 , 0.1) for the bottom-right panel, respectively.}
\end{figure}

\clearpage

\begin{figure}
\centering{
\includegraphics[scale=0.4]{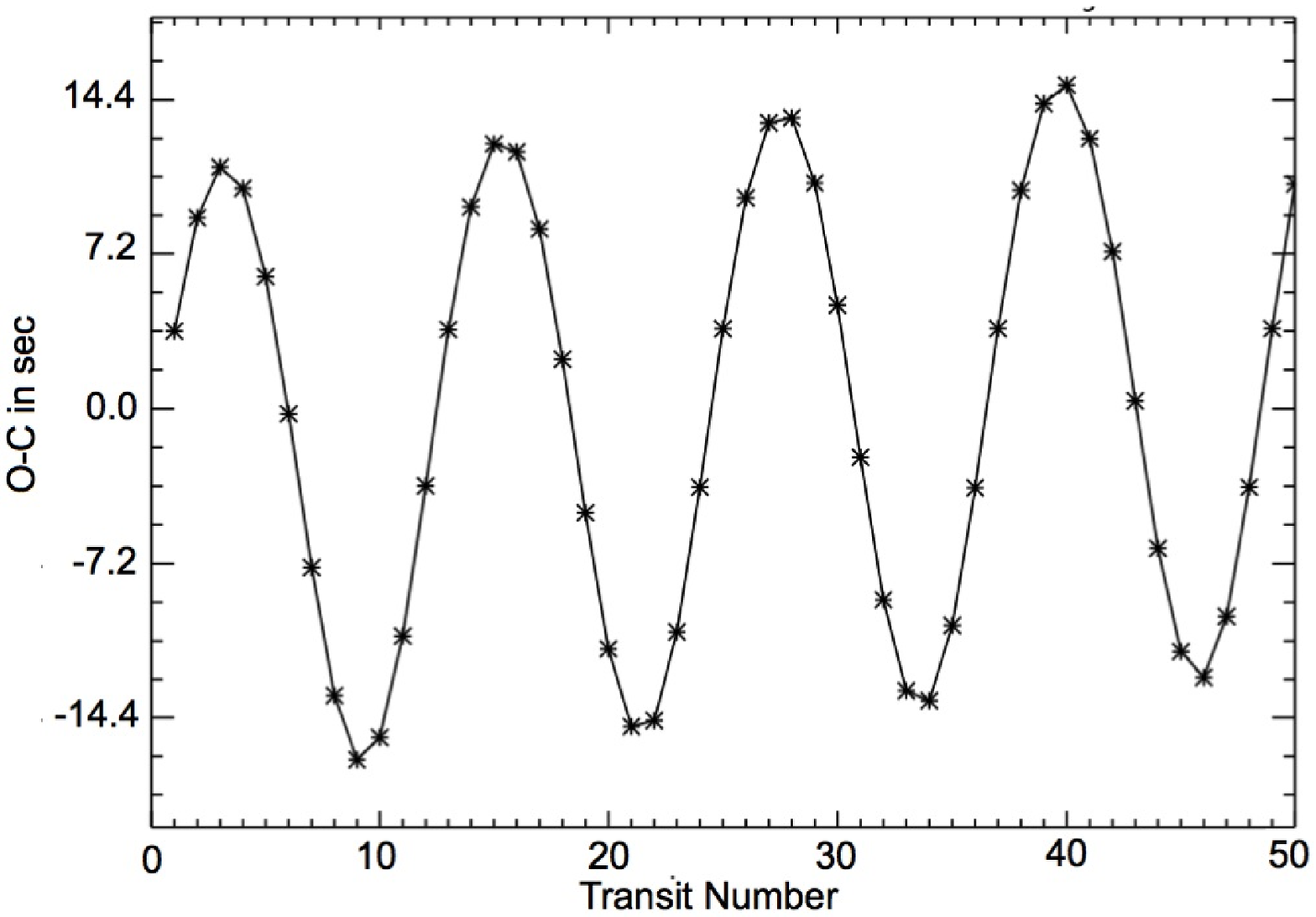}
\includegraphics[scale=0.4]{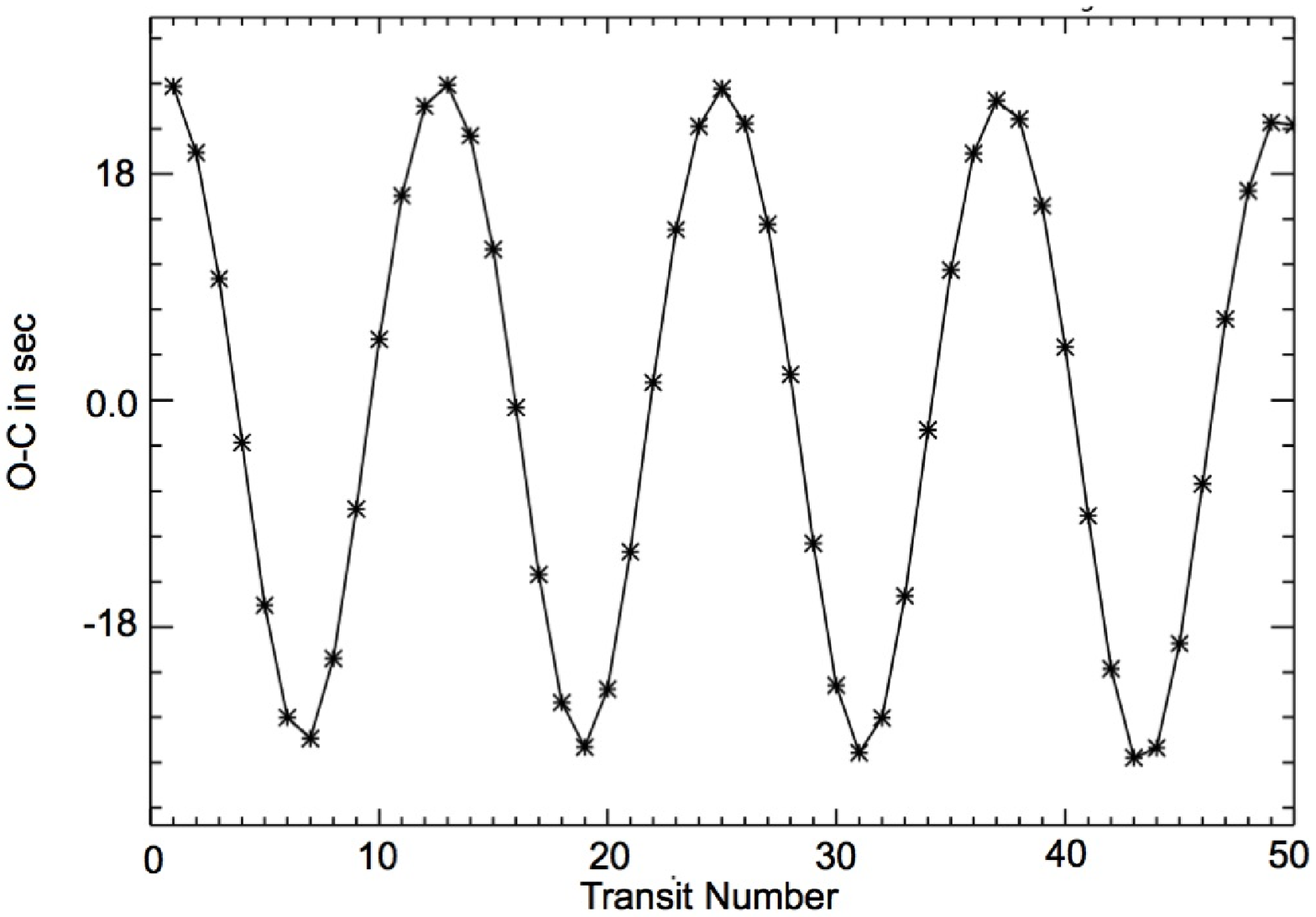}
\includegraphics[scale=0.4]{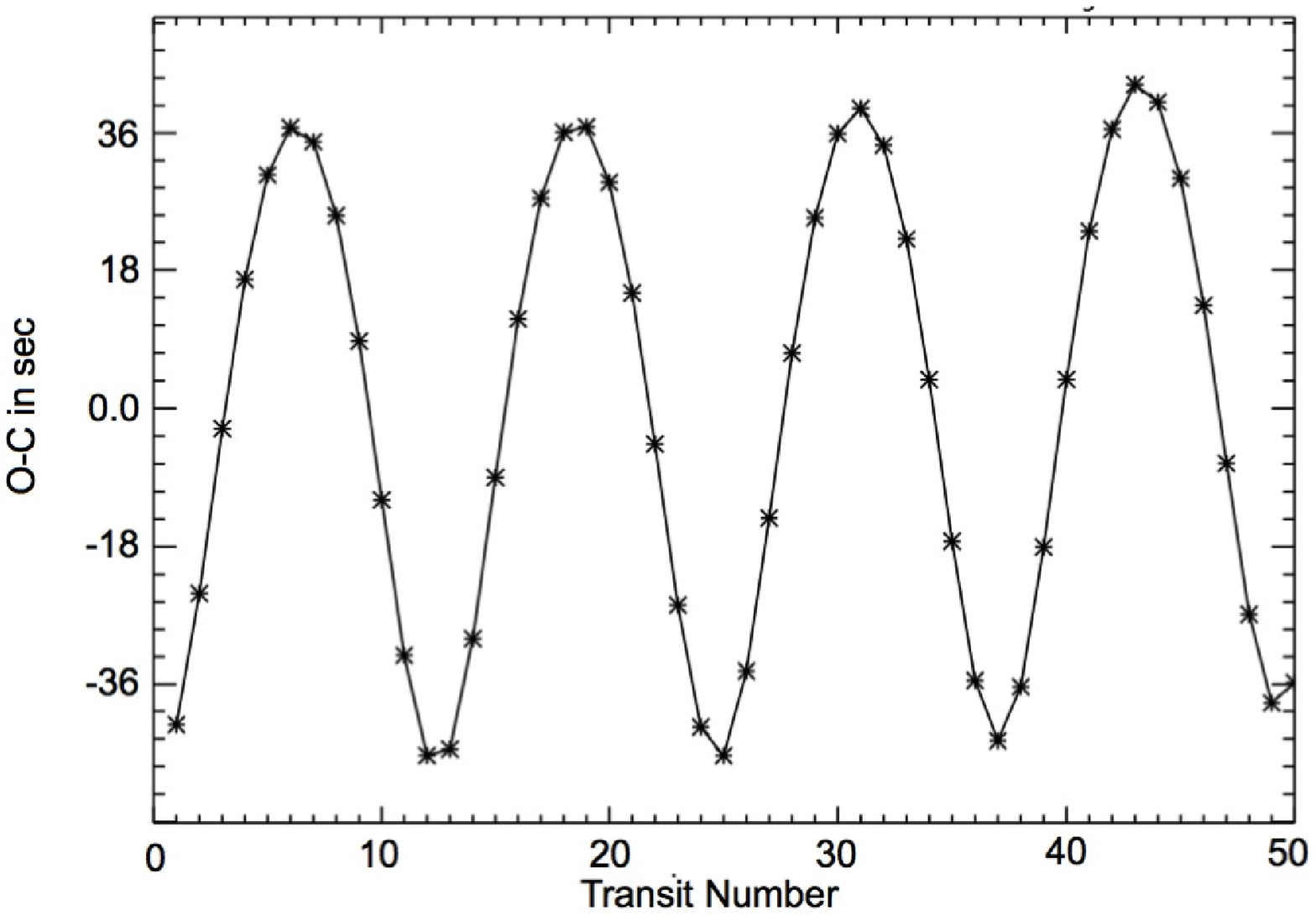}
\includegraphics[scale=0.4]{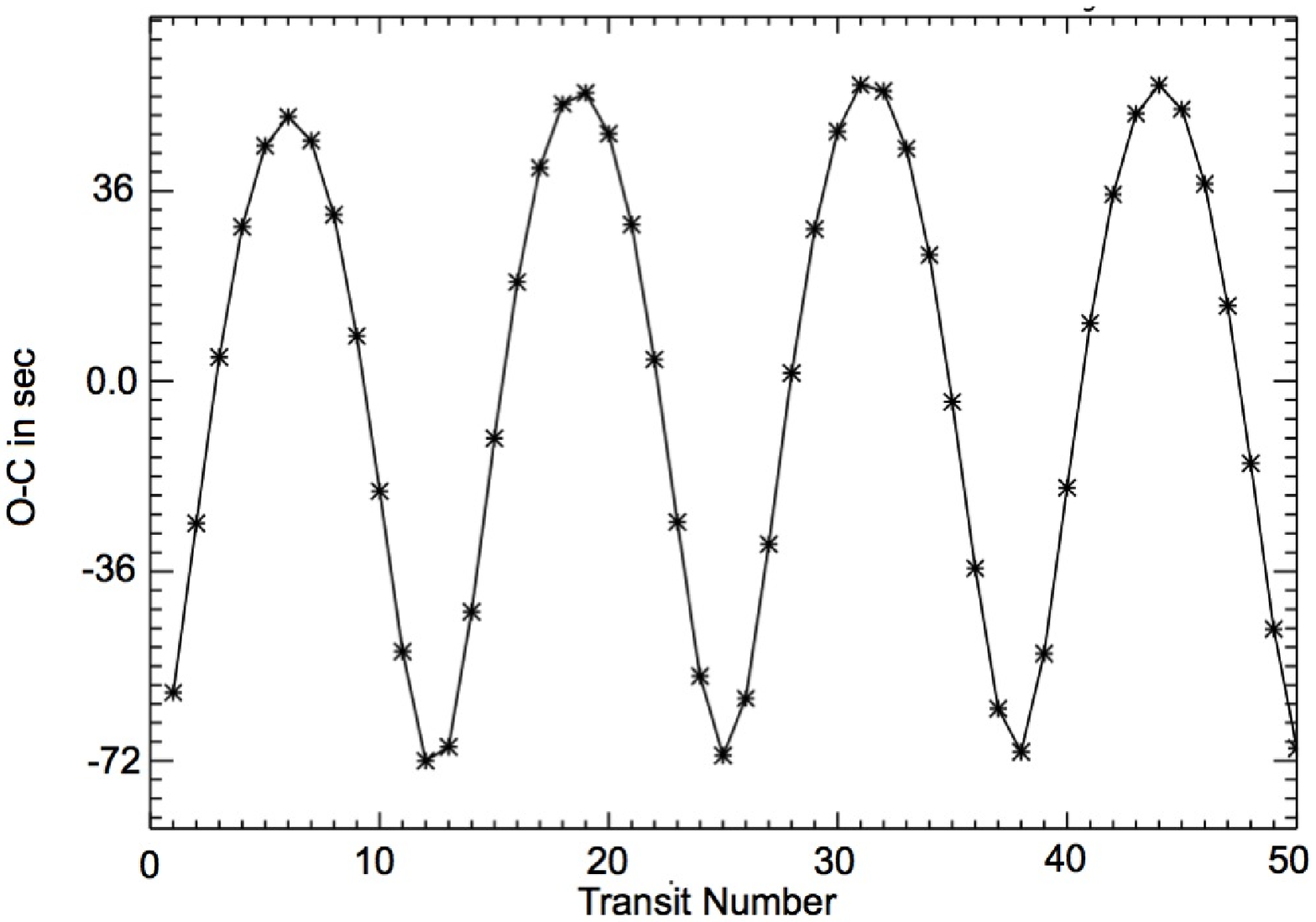}}
\caption{Graphs of the TTVs of a Jupiter-mass transiting planet due to an Earth-mass Trojan. 
The central star is Sun-like and the transiting planet is in a 3-day, circular orbit. The initial
orbital eccentricity of the Trojan planet was 0 (top-left), 0.05 (top-right), 0.1 (bottom-left), 
and 0.15 (bottom-right).
As shown here, the amplitude of TTVs increase by increasing the eccentricity of the Trojan planet
from $\sim 16$ s for zero eccentricity to $\sim 72$ s for an eccentricity of 0.1.}
\end{figure}

\clearpage

\begin{figure}
\centering{
\includegraphics[scale=0.4]{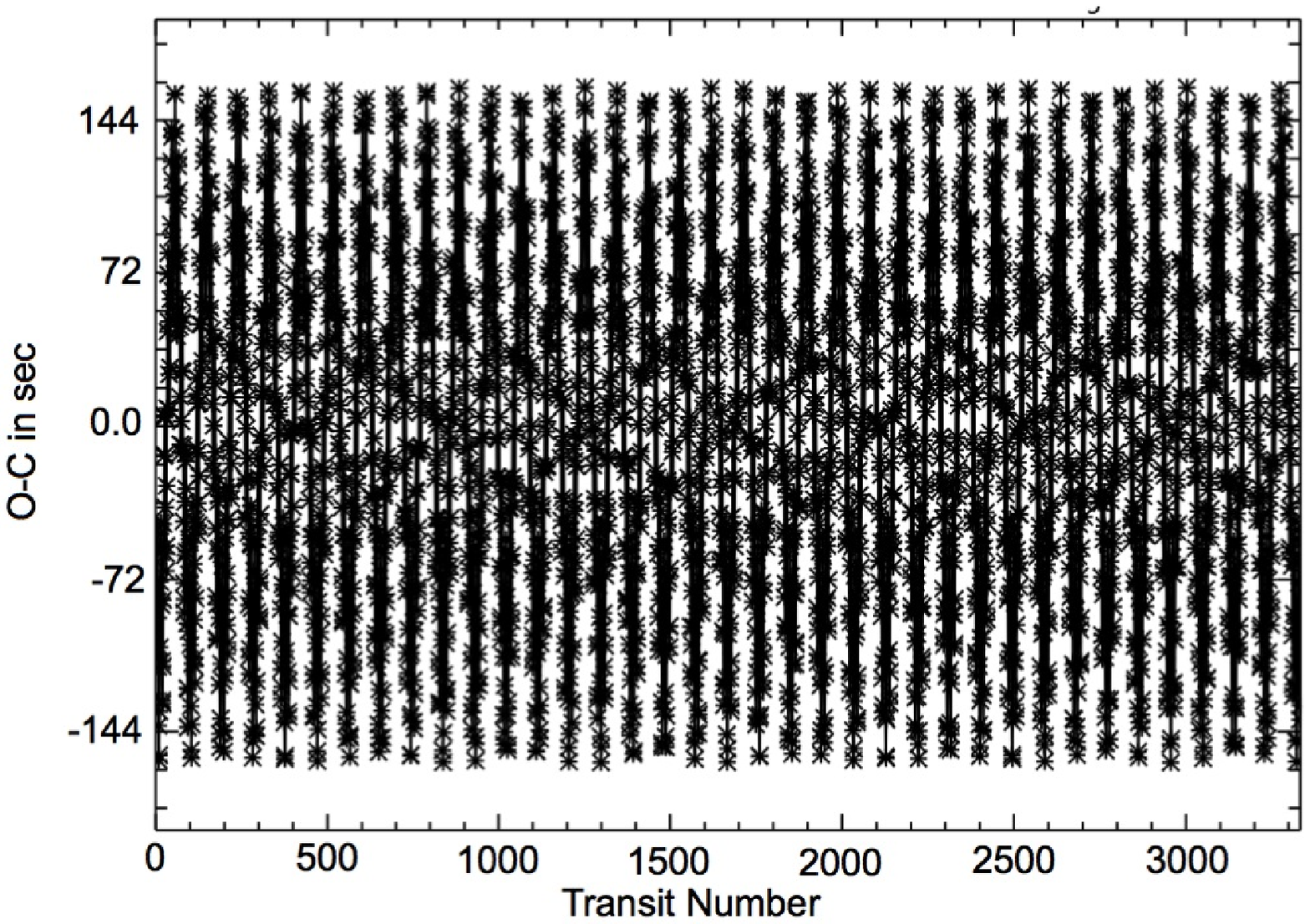}
\includegraphics[scale=0.4]{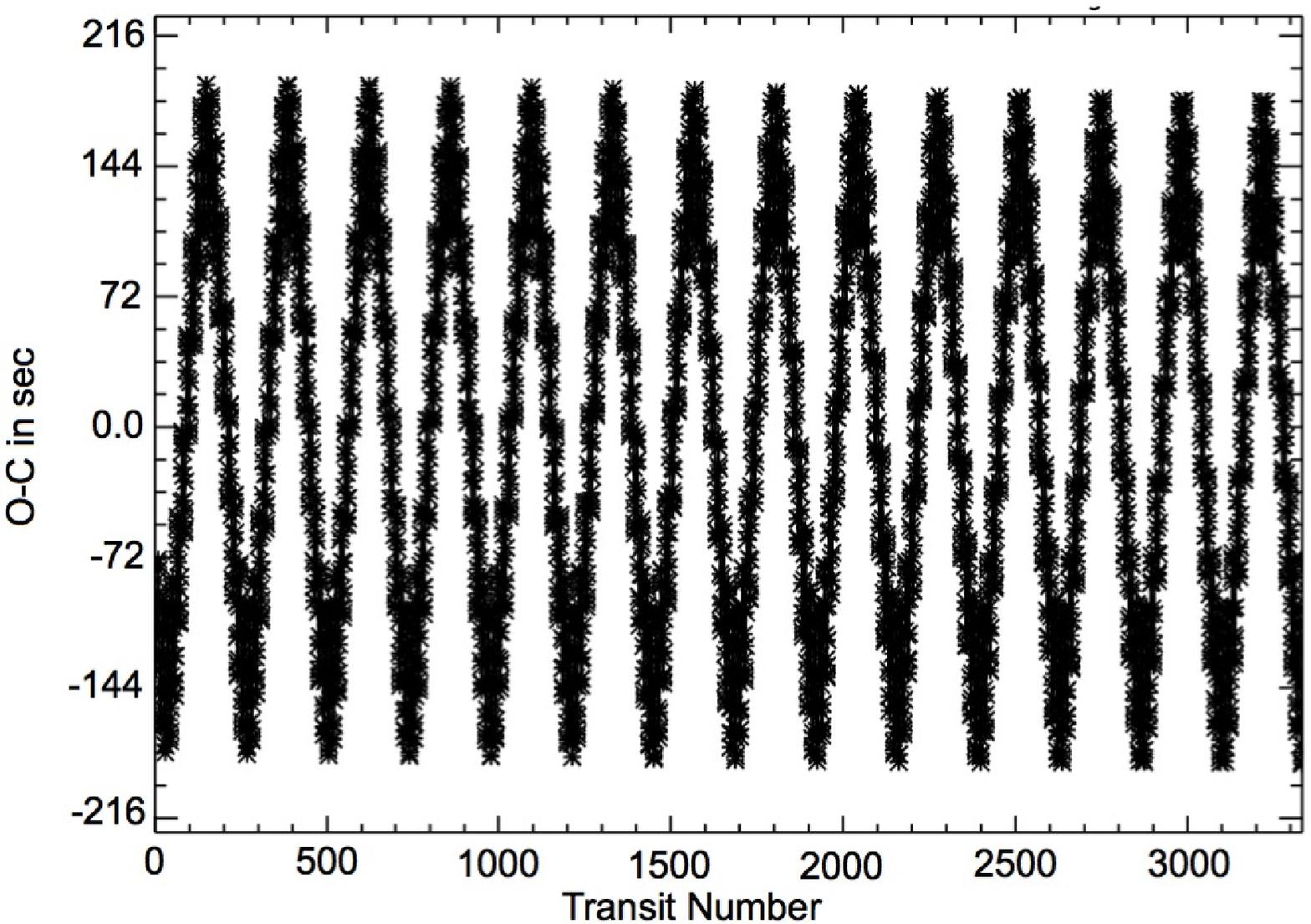}
\includegraphics[scale=0.4]{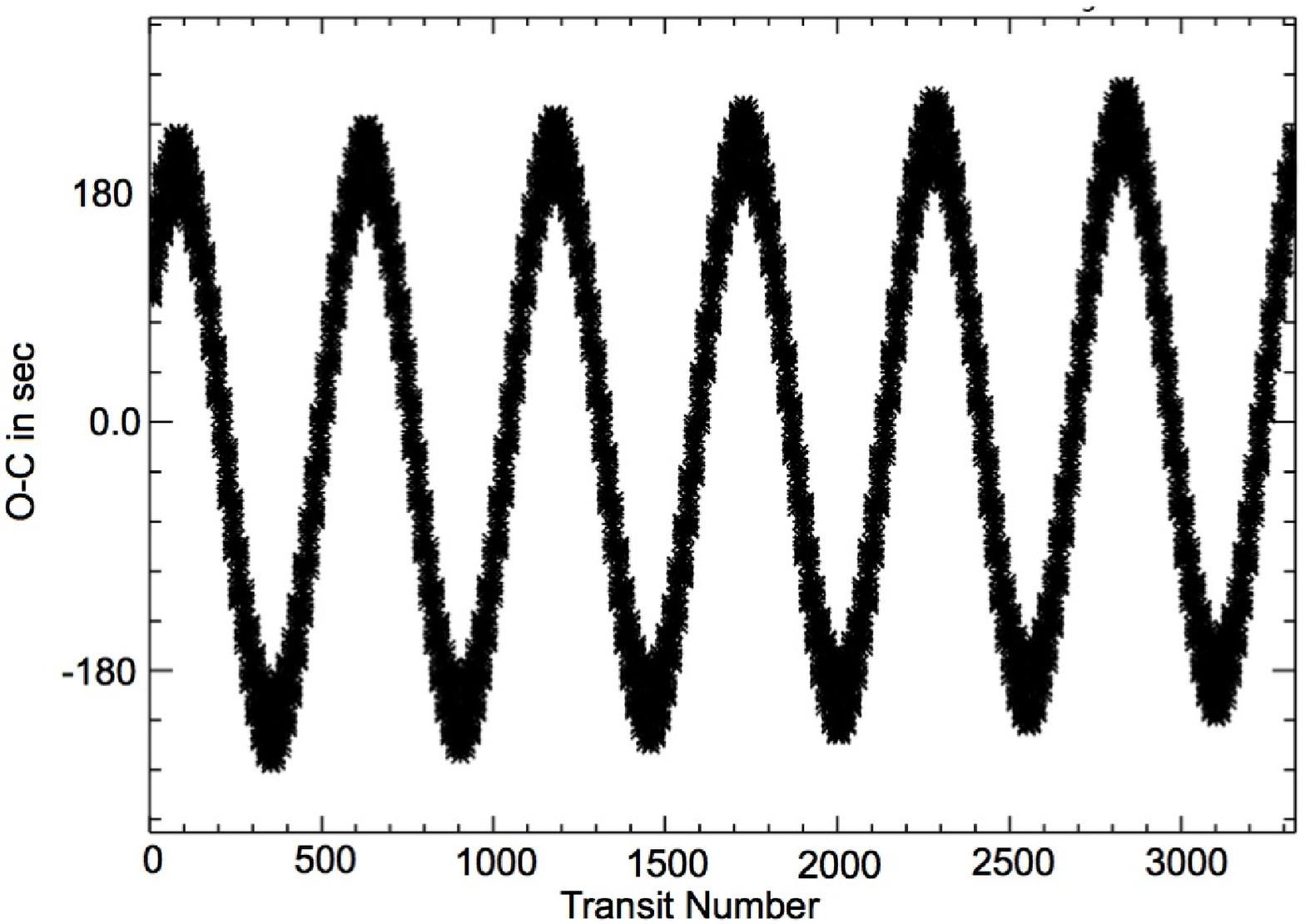}
\includegraphics[scale=0.4]{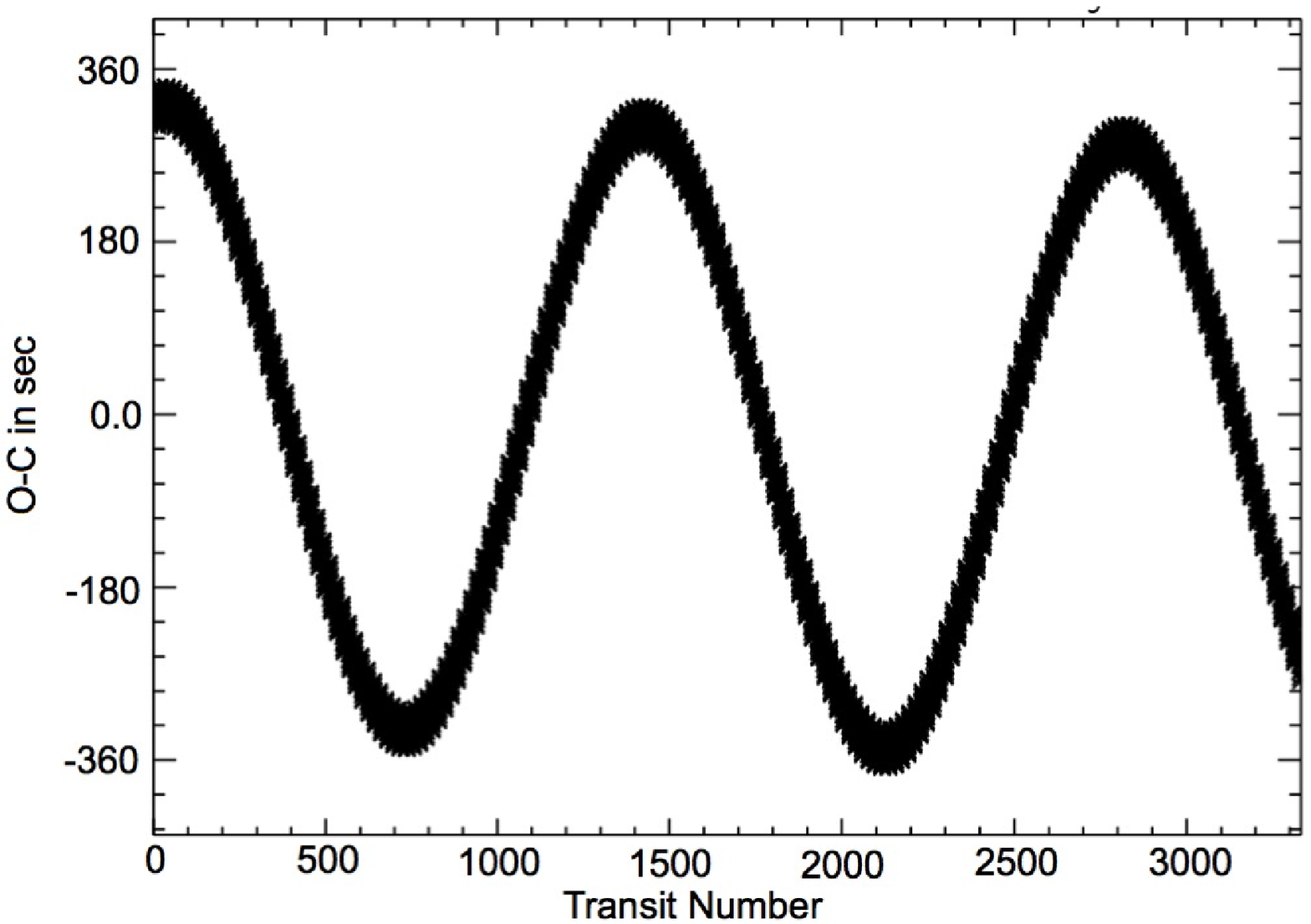}}
\caption{Graphs of the TTVs of a Jupiter-mass transiting planet in a 3-day, circular orbit due
to an Earth-mass perturber in a 1:1 MMR. The central star is solar-mass.
The perturber's orbital eccentricity is equal to 0.4 (top-left), 0.5 (top-right), 
0.6 (bottom-left), and 0.7 (bottom-right). Note that the time of integration is over three years. 
As shown by the bottom-right panel, the amplitude of TTV increases to approximately 6 minutes; 
a value that is within the range of the detected transit timing variations of the {\it Kepler}'s
planetary candidates.}
\end{figure}

\clearpage

\begin{figure}
\centering{
\includegraphics[scale=0.7]{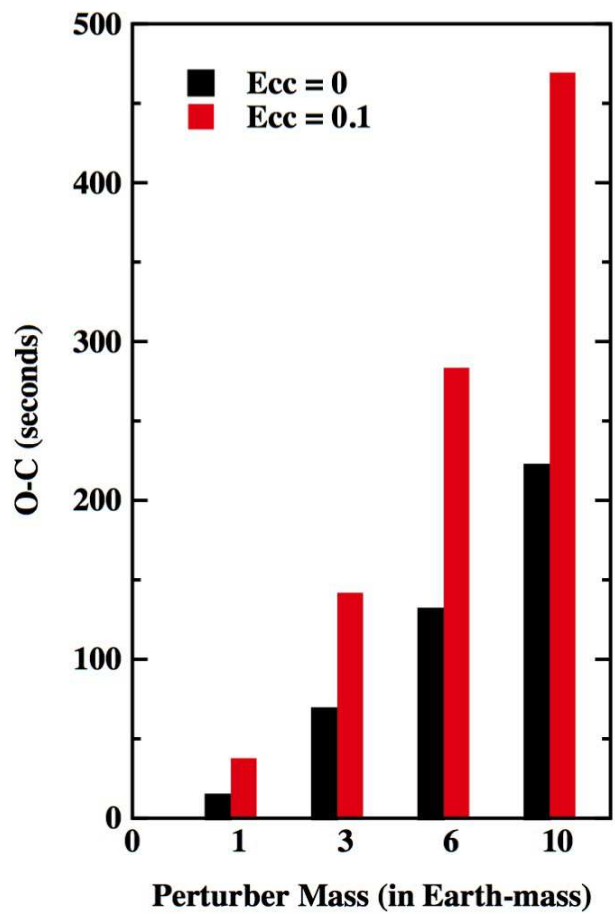}}
\caption{Graph of the maximum TTVs for different value of the mass of the Trojan perturber.
The central star is solar-type and the transiting planet is Jupiter mass in a 3-day, circular orbit.
The black and red colors correspond to different initial orbital eccentricity of the
Trojan planet. As expected, the amplitude of TTVs increase for larger values of the mass of this object.}
\end{figure}

\clearpage

\begin{figure}
\centering{
\includegraphics[scale=0.5]{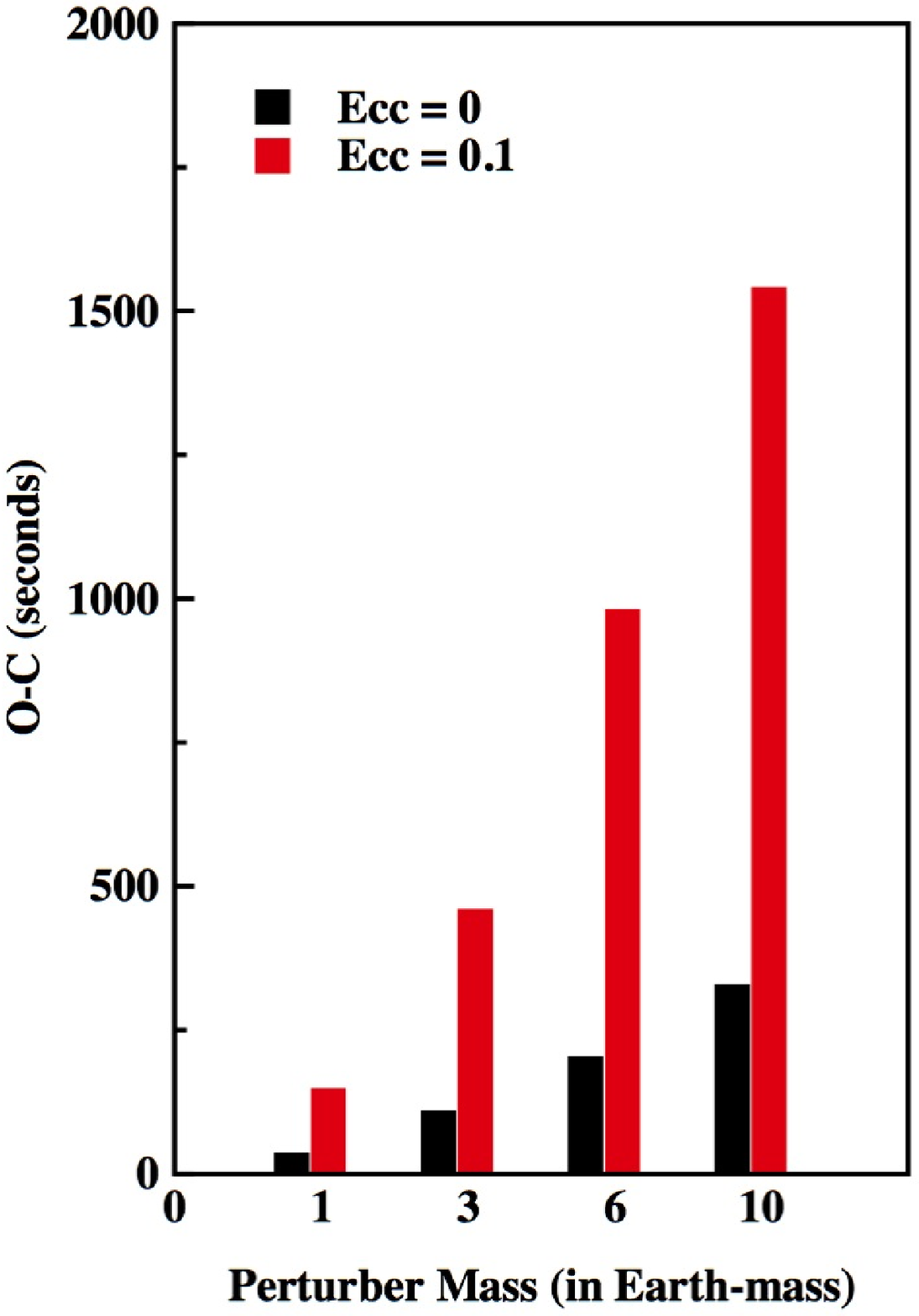}
\includegraphics[scale=0.5]{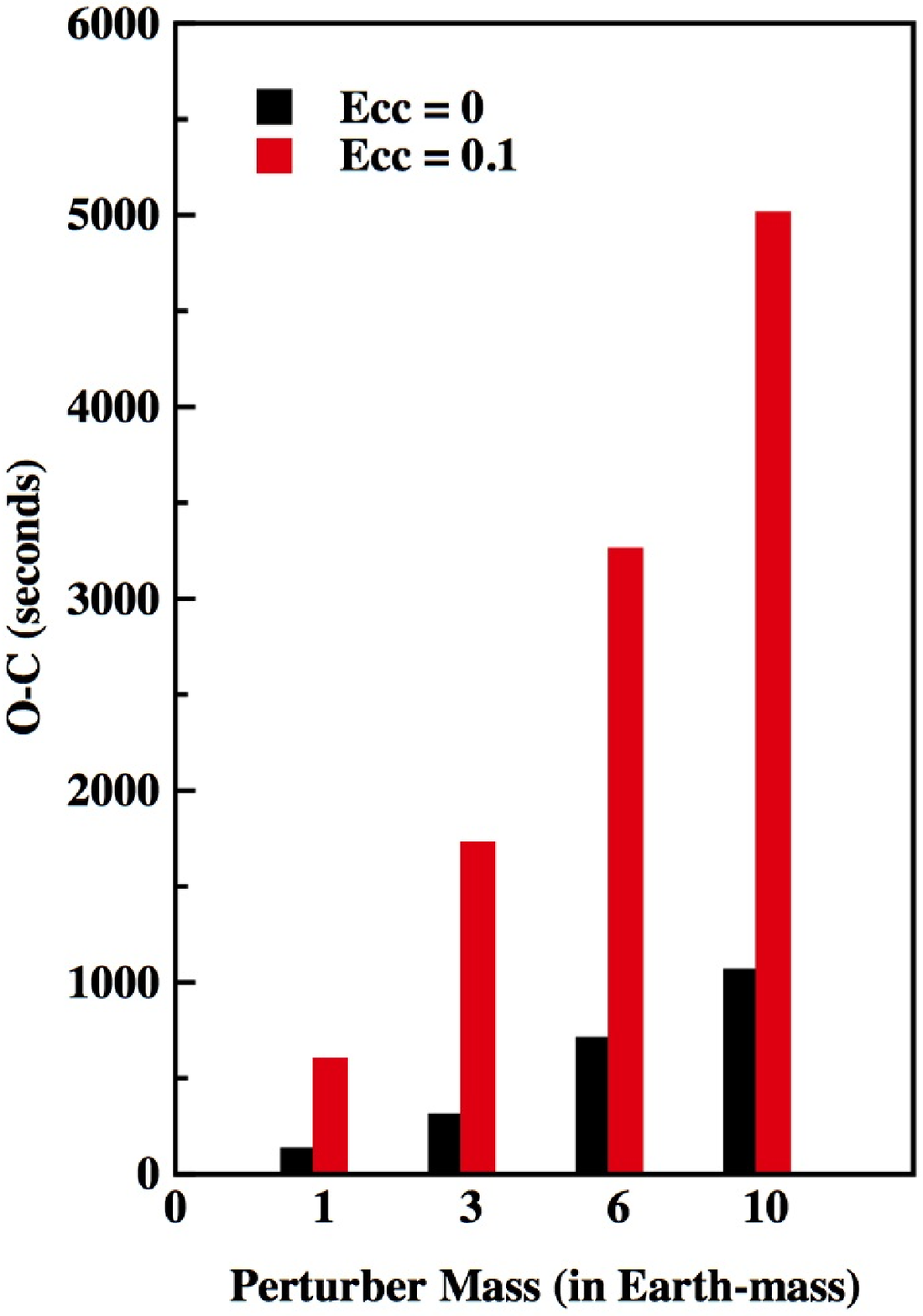}}
\caption{Graphs of the TTVs of a 0.1 Jupiter-mass (left) and 0.5 Jupiter-mass (right)
transiting planet due to a Trojan with a mass from 1 to 10 Earth-masses. The black and 
red colors correspond to different initial orbital eccentricity of the Trojan planet.
The transiting planet was initially in a circular orbit and the central star is Sun-like.}
\end{figure}

\clearpage

\begin{figure}
\centering{
\includegraphics[scale=0.4]{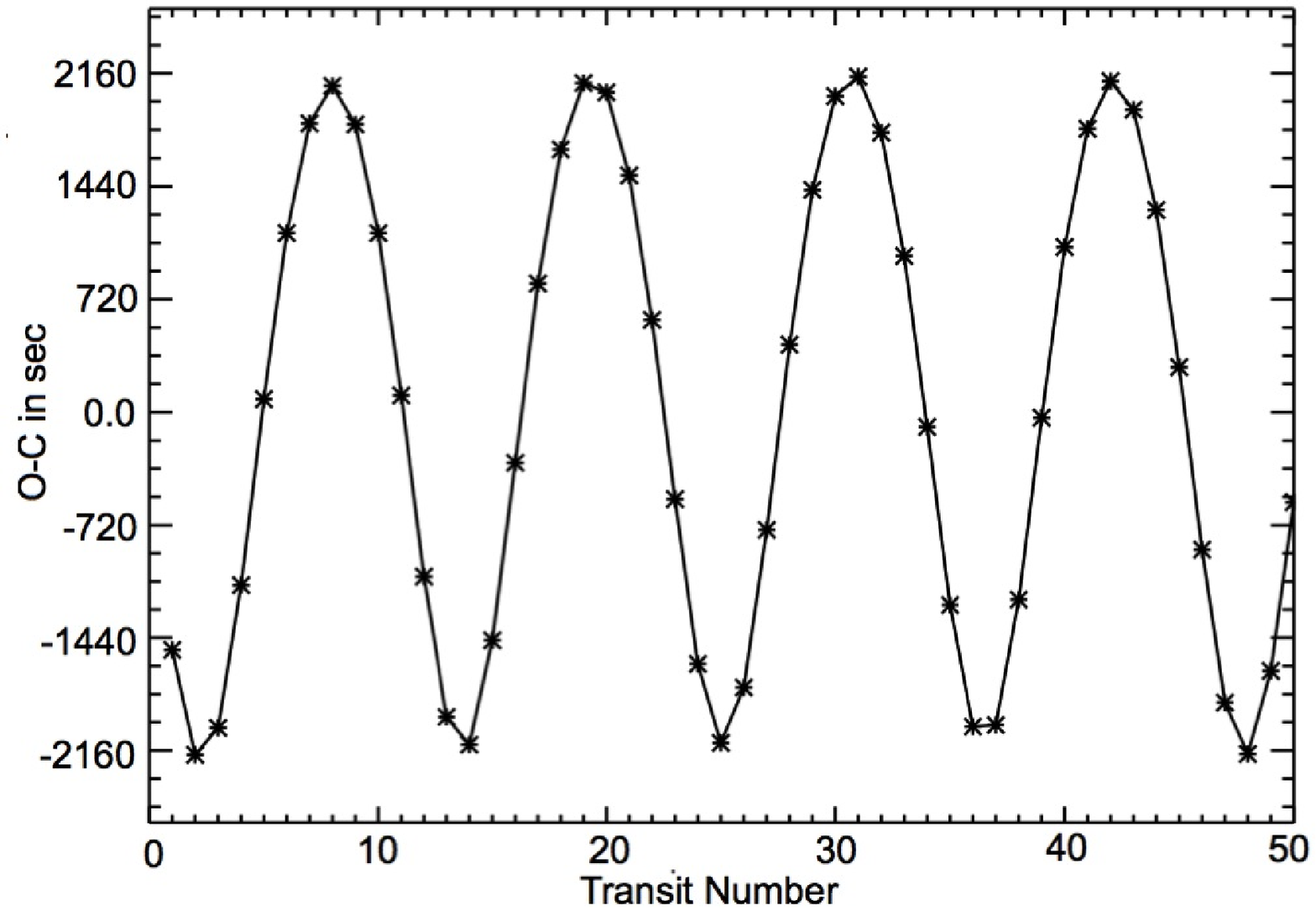}
\includegraphics[scale=0.4]{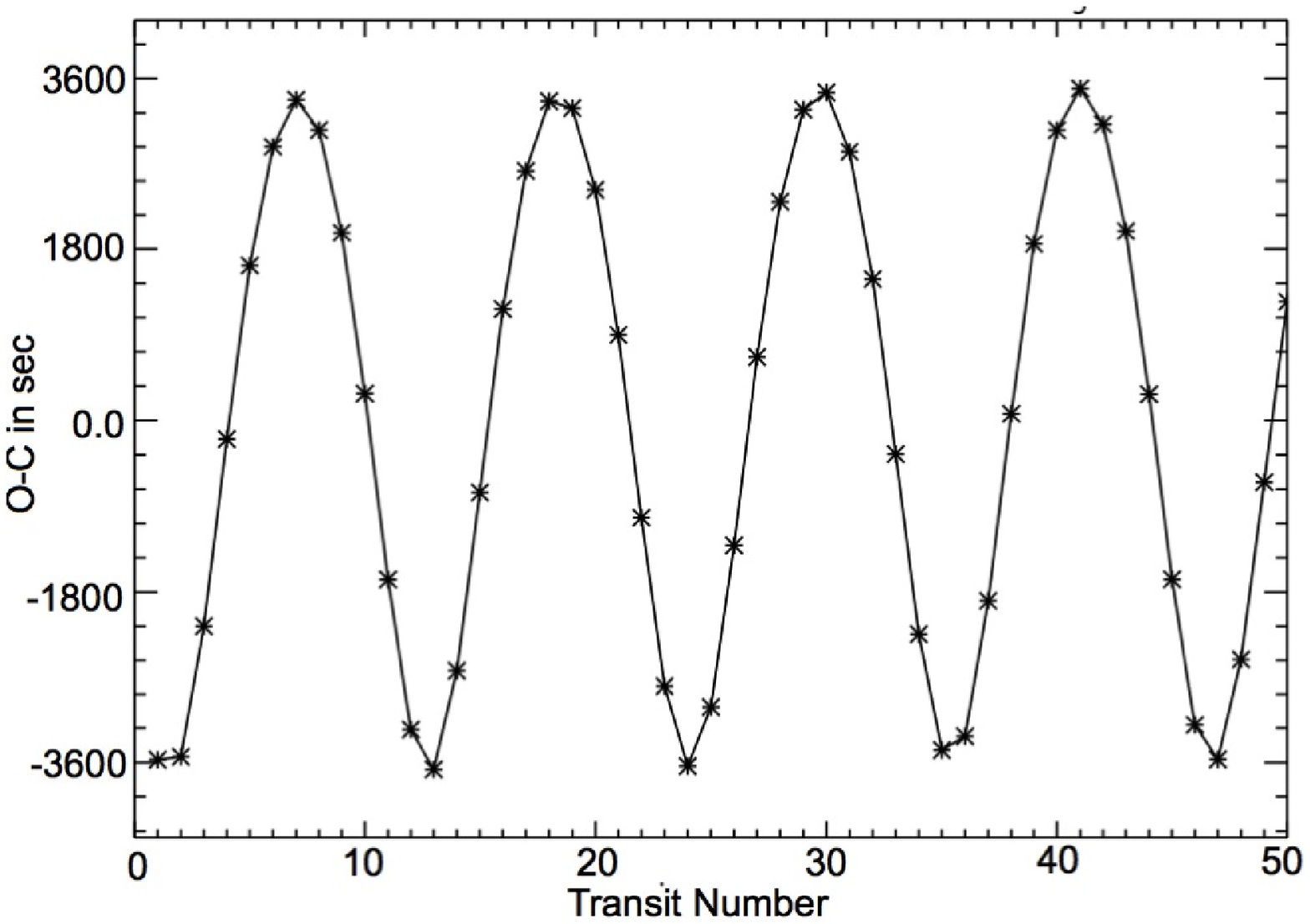}
\includegraphics[scale=0.4]{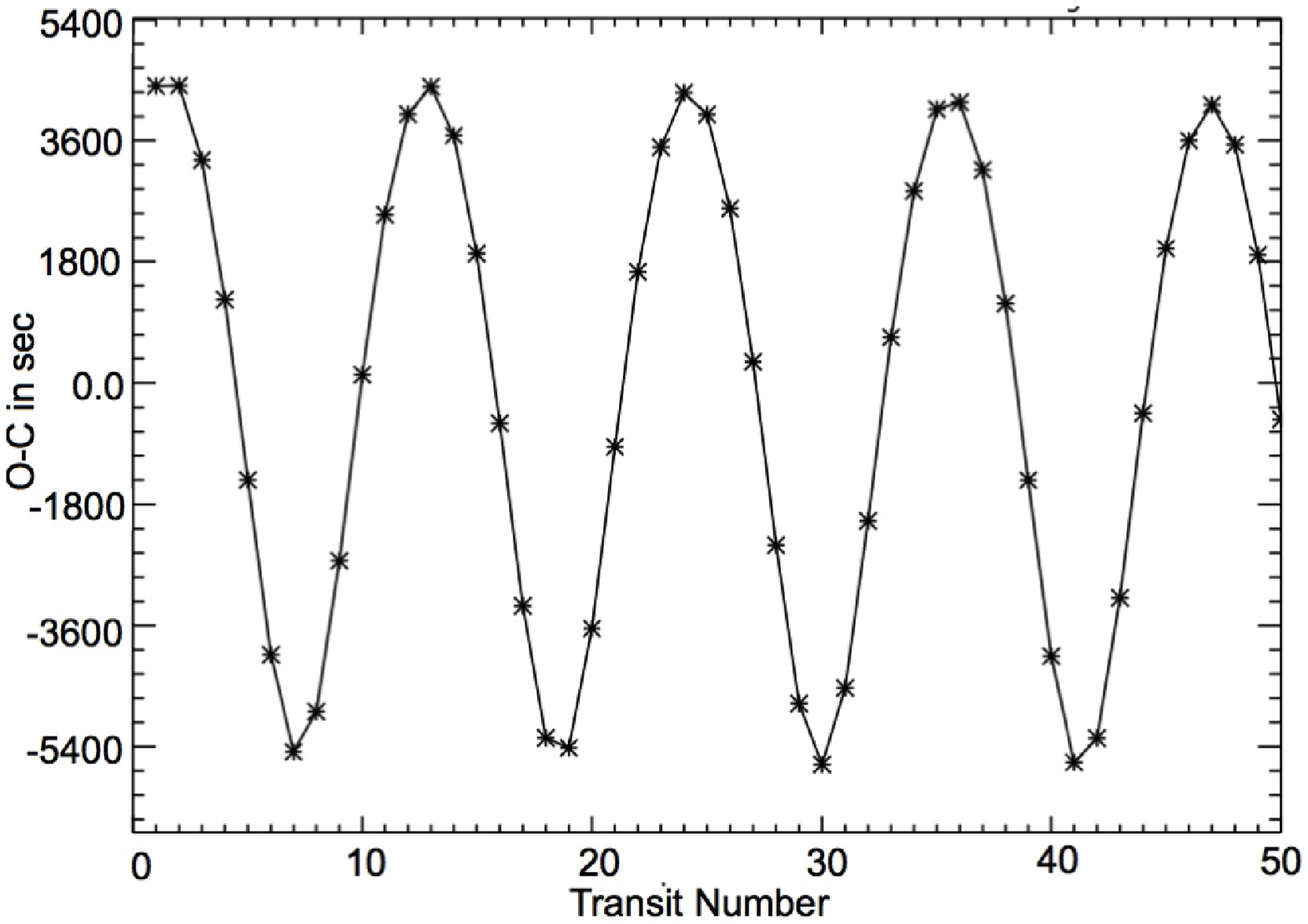}
\includegraphics[scale=0.4]{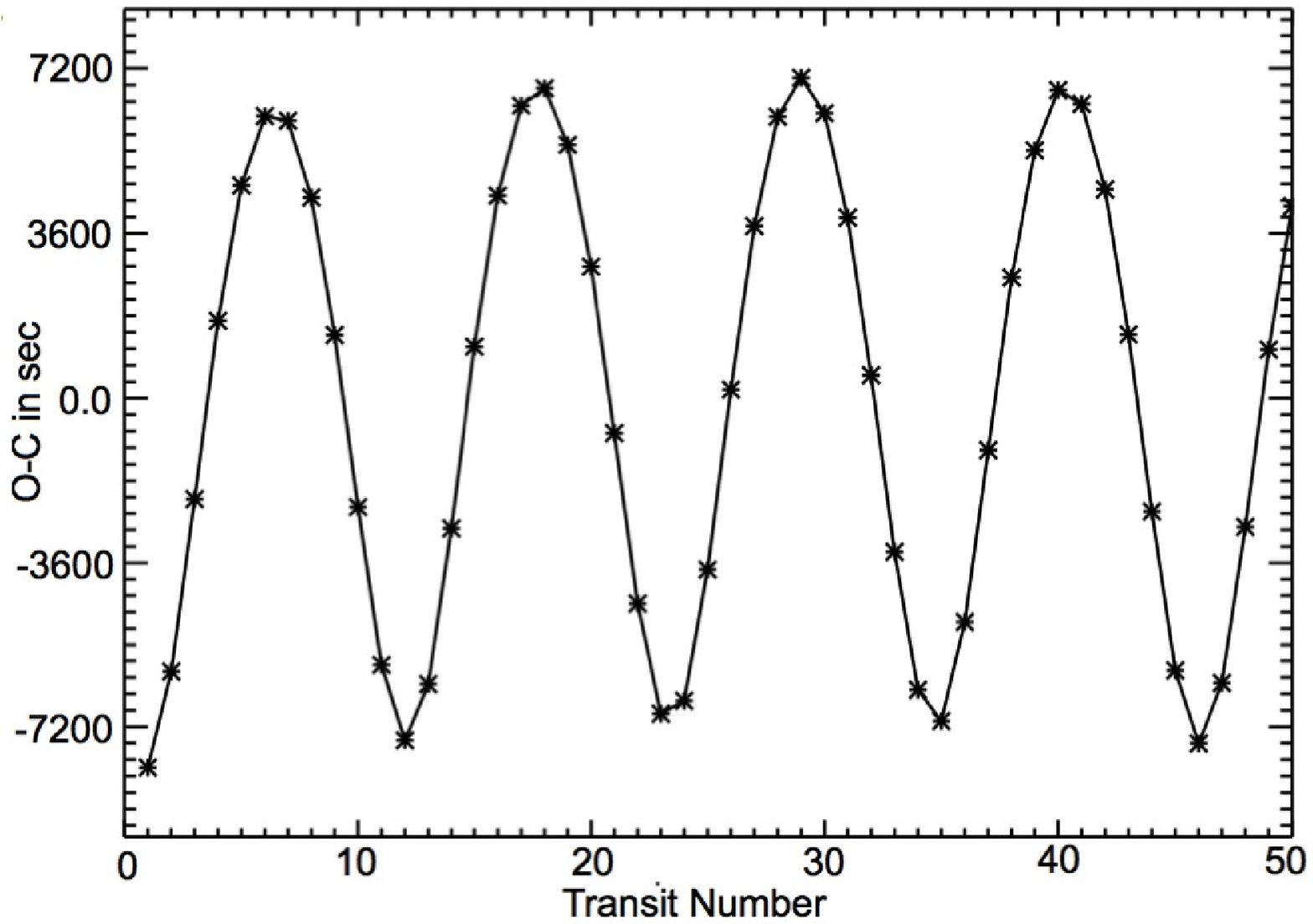}}
\caption{Graphs of the amplitudes of the TTVs of a 0.1 Jupiter-mass transiting
planet due to a 6 Earth-masses Trojan perturber. The transiting planet is in a circular
orbit and the central star is Sun-like. From top to bottom and clock-wise, the graphs correspond to the 
transiting planet in a 3-day (top-left), 5-day (top-right), 7-day (bottom-left), 
and 10-day (bottom-right) orbits.}
\end{figure}

\clearpage

\begin{figure}
\centering{
\includegraphics[scale=0.5]{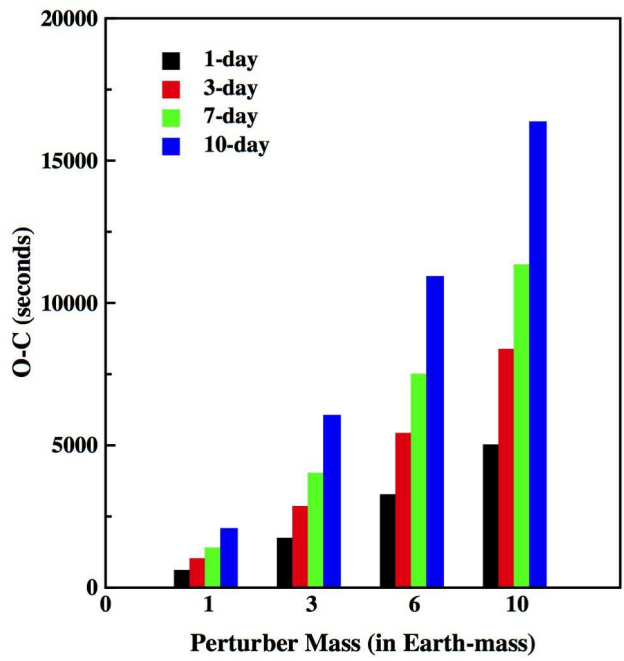}}
\caption{Graph of the maximum value of the TTV of a 0.1 Jupiter-mass transiting
planet due to a Trojan perturber with a mass from 1 to 10 Earth-masses. 
The transiting planet is in a circular orbit and the central star is Sun-like.
The colors corresponds to different orbital period of the transiting 
planet.} 
\end{figure}

\clearpage

\section{Appendix}

In general, determining the actual dynamical state of an object using MEGNO maps requires long-term
integrations. This is due to the fact that the detection of chaos with MEGNO 
depends strongly on the length of the overall integration time, and as a result, integrating for long times
become crucial for obtaining the true dynamical picture of the underlying structure of the phase space.
Long-term integrations are particularly important when attempting to detect chaotic (secular and mean-motion)
orbital resonances. To determine to what extent the MEGNO maps of our systems represent the true dynamical 
stability of the perturber, we considered a very narrow range of orbital eccentricity, $e=0.7$,
(the black horizontal line in the top-left panel of figure 2 in the main text) 
and calculated a one-dimensional MEGNO over
the range of the perturber's semimajor axis, and for longer times. Figure 1 shows the results of three
different integration time spans, 100 years (top), 1000 years (middle), and 10,000 years (bottom). 
We considered three different integrations times in order to be able to study the evolution of the
MEGNO and determine the time beyond which the structure of the phase space will stay unchanged. For instance,
an inspection of the panels in this figure shows that the initial conditions at the point 0.0435 AU
at the top and middle panels are quasi-periodic approaching a MEGNO of 2. However, as the integrations
continue, the quasi-periodic nature of this point disappears and it becomes unstable. Our integrations
indicated that 10,000 years would be sufficiently long to capture a close-to-complete qualitative
picture of the dynamical state of the perturber. After this time, the maps maintained
their architecture and did not change, suggesting that their depicted orbital (ir)regularity 
is a true representation of the dynamical (in)stability of the system. Figures 2 and 3 show 
samples of the results as examples of the quasi-periodic (stable) and chaotic (in this case, unstable) orbits.


\begin{figure}
\centering{
\includegraphics[scale=0.7]{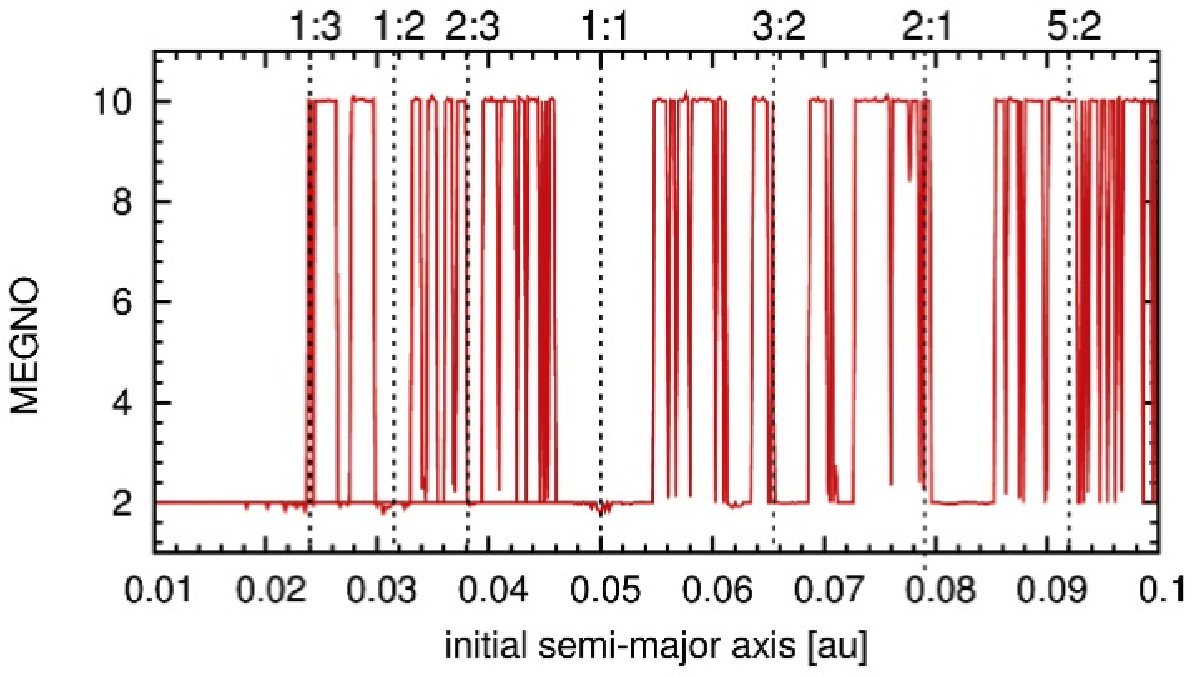}
\vskip 5pt
\includegraphics[scale=0.7]{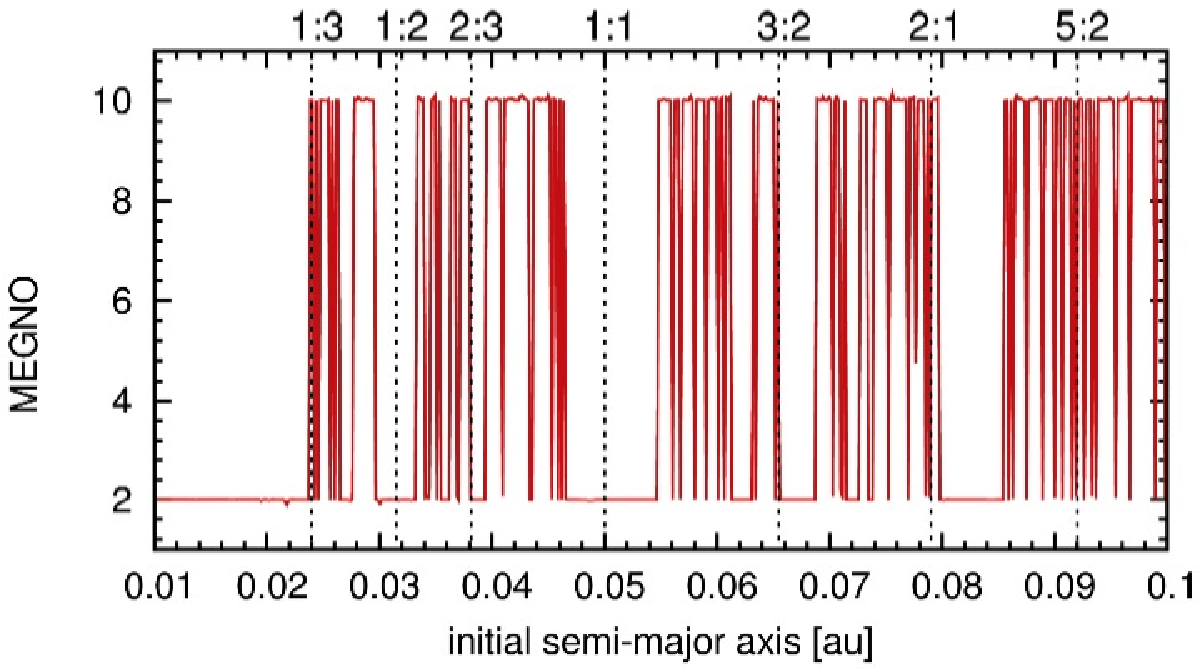}
\vskip 5pt
\includegraphics[scale=0.7]{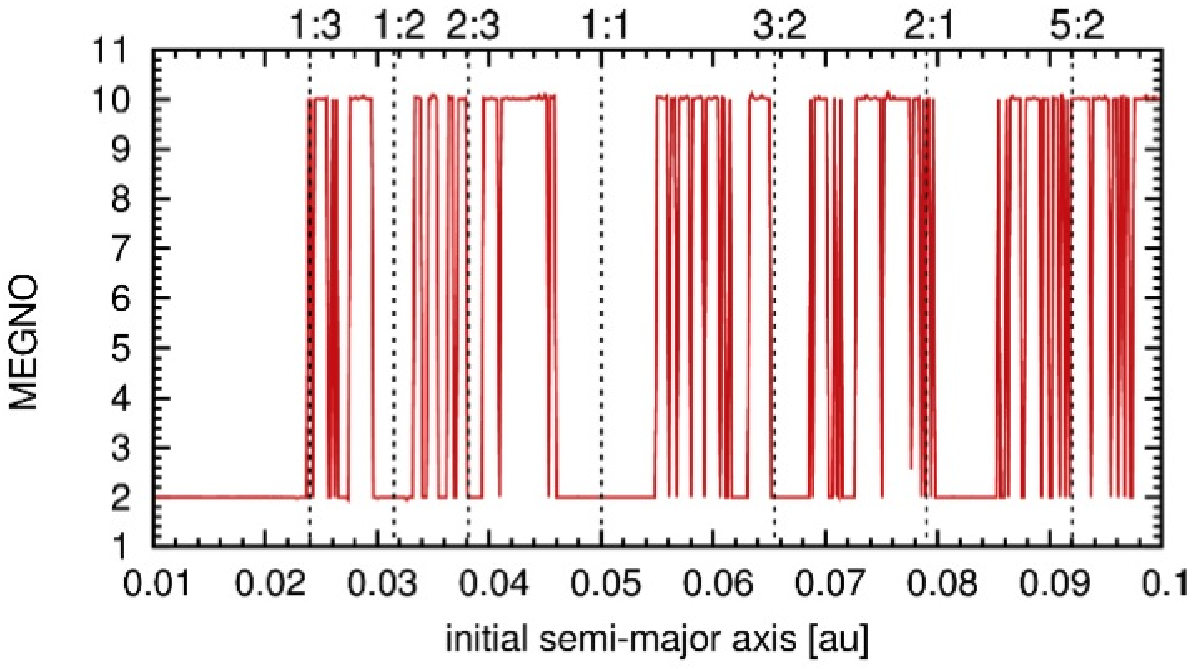}}
\caption{One-dimensional MEGNO maps for an Earth-mass perturber in the system of figure 2
in the main text. The initial 
eccentricity of the perturber was 0.7. See the black horizontal  line on the top-left panel of figure 2.
The panels show the MEGNO for different times of integration; 100 years (top), 1000 years (middle),
and 10,000 years (bottom). As shown here, when the integration reaches long times, the MEGNO maps
do not change indicating that their regions of regular orbits represent actual stability of the
perturbing planet.}
\end{figure}

The initial values of the eccentricity and semimajor axis
of the perturber for these figures are shown by black crosses in the top-left panel of figure 2
in the main text.
Figure 2 shows details of the time evolution of MEGNO for a quasi-periodic (stable) initial condition 
(the black cross from the low-MEGNO region around the 1:1 MMR). As shown here, the value of MEGNO 
approaches 2.0 asymptotically and stays in that level for the remainder of the integrations. 
The perturber corresponding to figure 3, however,
shows an entirely different behavior. The integrations of the orbit of this object, whose initial semimajor 
axis was chosen from the high-MEGNO part of the upper left panel of figure 2 in the main text, 
indicate that the
eccentricity of its orbit rapidly increases causing the object to collide with the central body in a
short time. As expected, the magnitude of the MEGNO of this body also reaches very high values 
during this time.


\begin{figure}
\centering{
\includegraphics[scale=0.45]{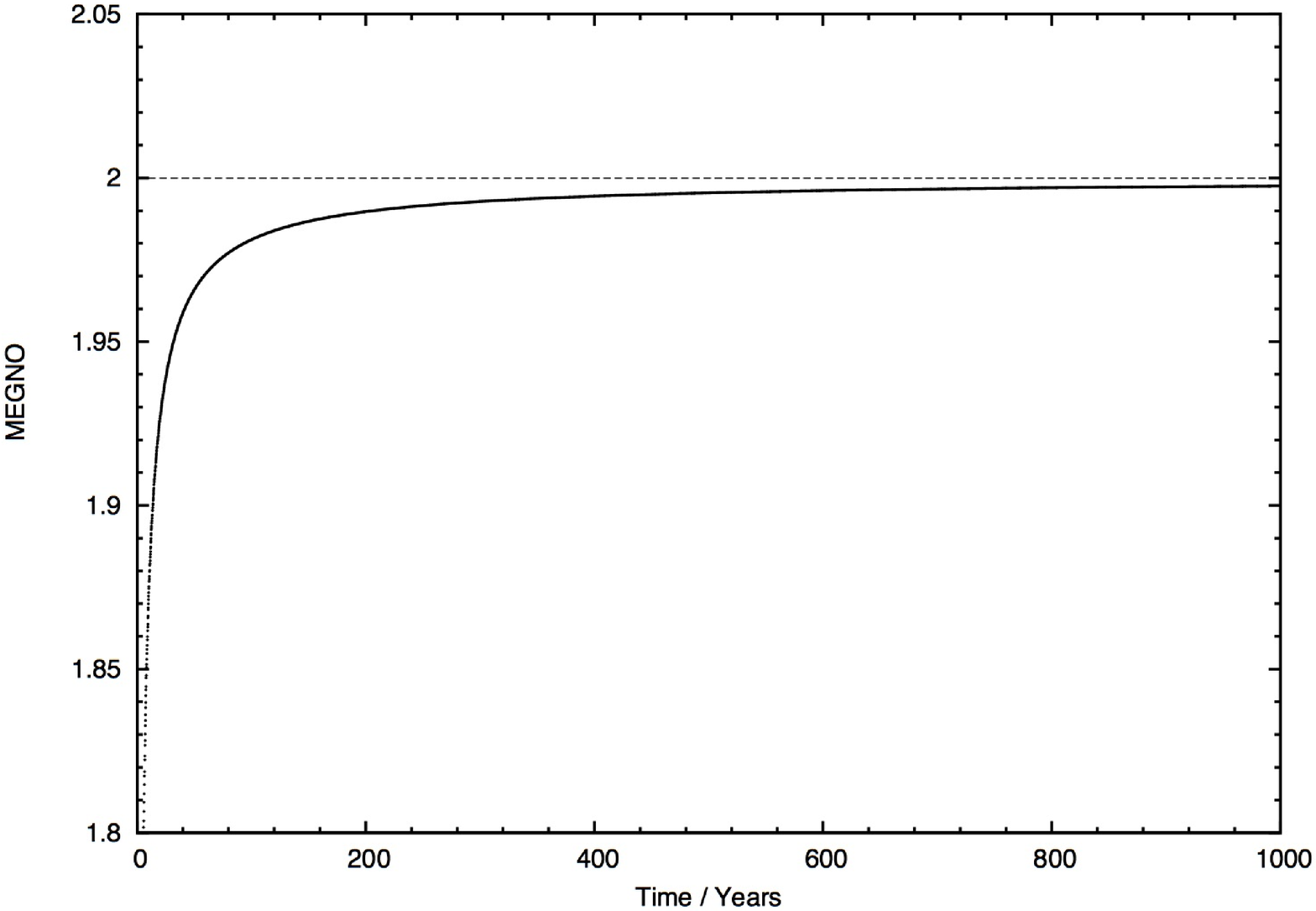}}
\caption{The graph of the value of the MEGNO of an Earth-mass perturber shown by the black cross in the 
low-MEGNO region of the 1:1 MMR in the upper left panel of figure 2 in the main text. 
As shown here, at the beginning of the integration,
the value of the MEGNO of the planet rapidly approaches 2 and stays in that level for the rest
of the integration indicating that the orbit of the planet is long--term stable.}
\end{figure}


\begin{figure}
\centering{
\includegraphics[scale=0.35,angle=-90]{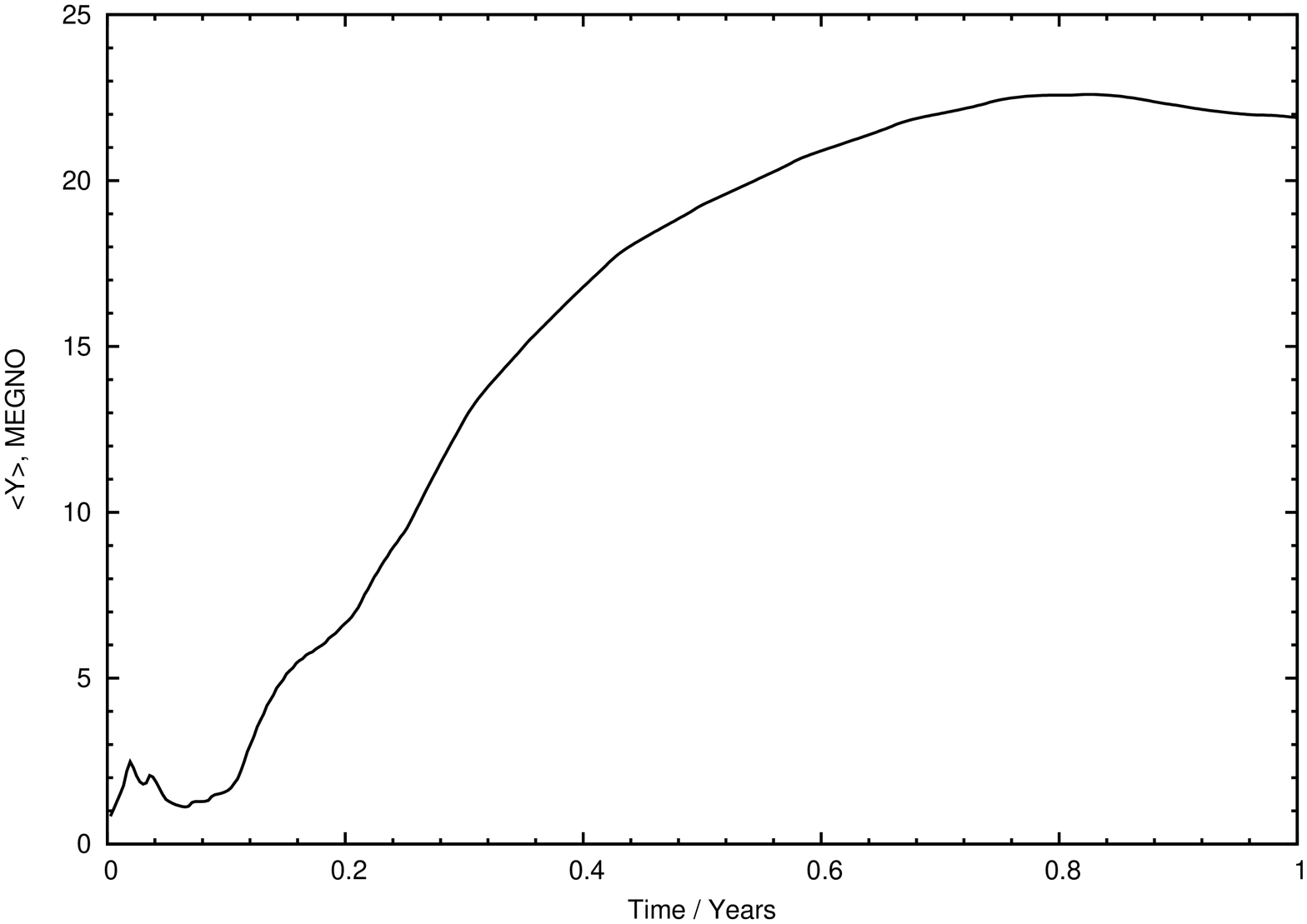}
\vskip 15pt
\includegraphics[scale=0.35,angle=-90]{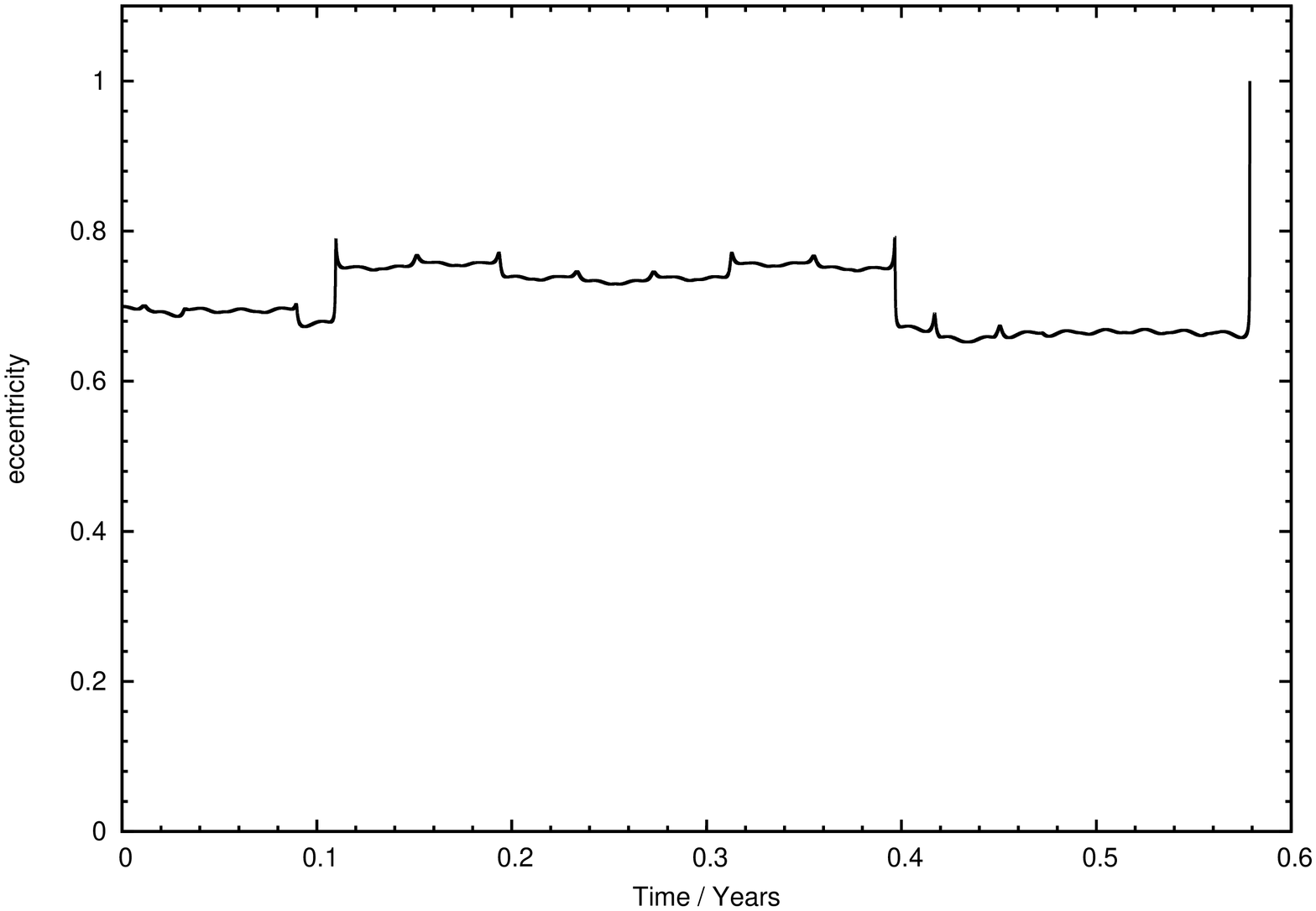}
\vskip 15pt
\includegraphics[scale=0.35,angle=-90]{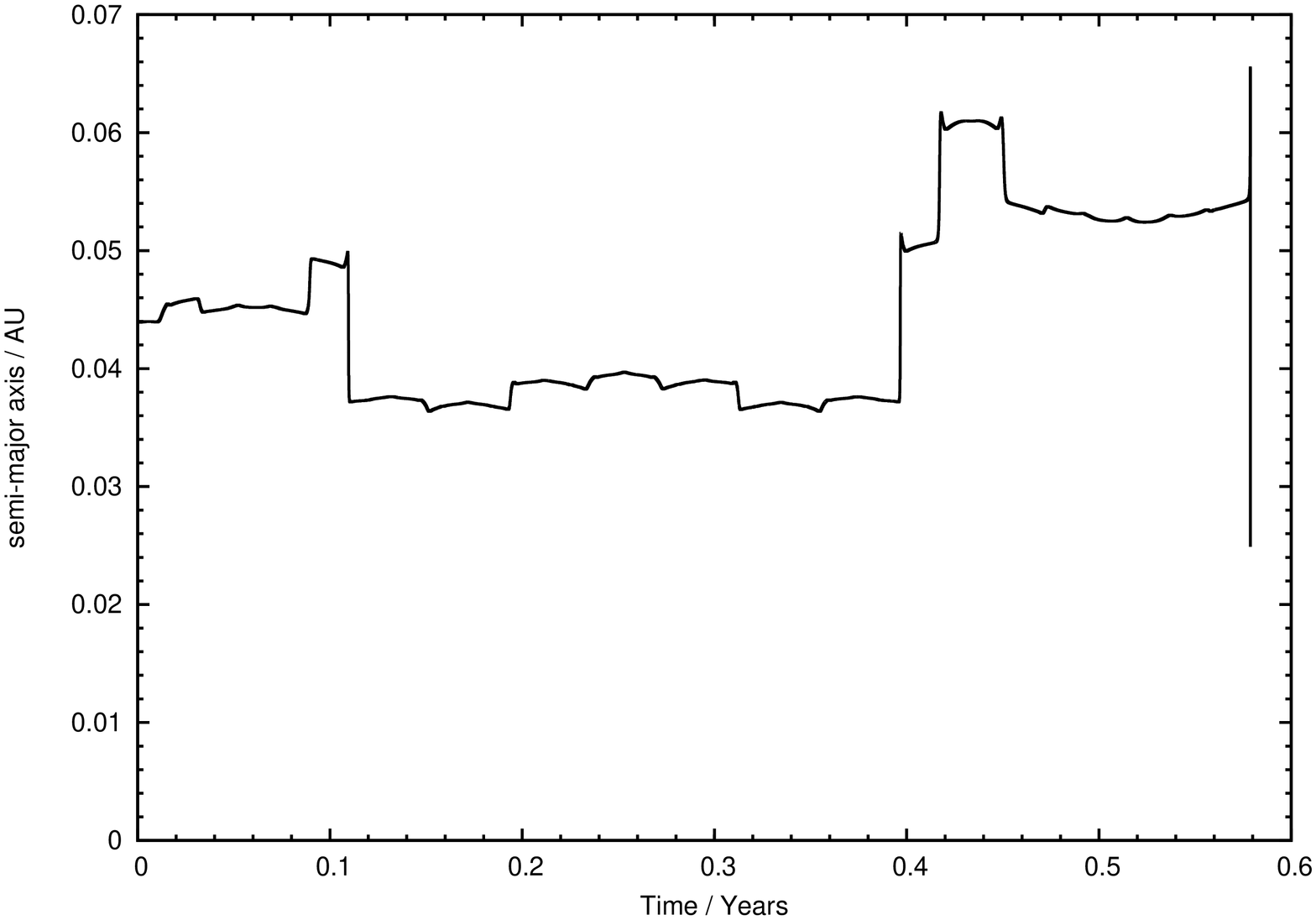}}
\caption{Graphs of the MEGNO, eccentricity and semimajor axis of the Earth-mass perturber
indicated by the left black cross in the top-left panel of figure 2 in the main text. 
As shown here, the planet's orbit
becomes unstable shortly after the start of the integrations when it collides with the central star.
As expected, the magnitude of the MEGNO of the planet raises to very high values during this time.}
\end{figure}

\end{document}